\documentclass[a4paper,fleqn]{cas-sc}
\usepackage[numbers]{natbib}
\usepackage{graphicx}
\usepackage{tikz}
\usepackage{tikz-network}
\usetikzlibrary{3d} 
\usetikzlibrary{backgrounds} 
\usetikzlibrary{shapes, arrows, positioning}
\usepackage{amsmath}
\usepackage{amssymb}
\usepackage{mathtools,xparse}
\usepackage{textcomp}
\usepackage[official]{eurosym}
\usepackage{xcolor} 
\usepackage{lipsum} 
\usepackage{siunitx}
\usepackage{placeins}

\usepackage{algorithm}
\usepackage{algpseudocode}
\usepackage{tabularx} 
\usepackage{cleveref}
\usepackage{float}
\usetikzlibrary{matrix}
\usepackage{subcaption} 
\usetikzlibrary{spy}
\usetikzlibrary{decorations.pathreplacing}
\usepackage{fontawesome5}

\newcommand{\xr}[1]{x_{ijbt}^{\text{r}}}
\newcommand{\xp}[1]{x_{ijbt}^{\text{p}}}
\newcommand{\xc}[1]{x_{iibt}^{\text{c}}}
\newcommand{\xvtg}[1]{x_{iibt}^{\text{V2G}}}

\DeclareMathOperator*{\minimize}{minimize}
\usetikzlibrary{
    shapes.geometric,
    arrows.meta,
    positioning,
    fit,
    backgrounds,
    calc,
    patterns,
    decorations.pathreplacing
}

\definecolor{encoderblue}{RGB}{70, 130, 180}
\definecolor{decoderred}{RGB}{180, 70, 70}
\definecolor{inputgreen}{RGB}{34, 139, 34}
\definecolor{outputpurple}{RGB}{128, 0, 128}
\definecolor{rewardyellow}{RGB}{160, 140, 30}
\definecolor{lightblue}{RGB}{220, 235, 250}
\definecolor{lightred}{RGB}{255, 230, 230}
\definecolor{lightgreen}{RGB}{220, 250, 220}
\definecolor{lightyellow}{RGB}{255, 250, 210}
\definecolor{lightpurple}{RGB}{240, 220, 255}
\definecolor{edgeorange}{RGB}{210, 130, 40}
\definecolor{lightorange}{RGB}{255, 235, 200}
\definecolor{nodecyan}{RGB}{50, 160, 160}
\definecolor{lightcyan}{RGB}{200, 240, 240}
\definecolor{policymagenta}{RGB}{180, 50, 120}

\def\tsc#1{\csdef{#1}{\textsc{\lowercase{#1}}\xspace}}
\tsc{WGM}
\tsc{QE}
\tsc{EP}
\tsc{PMS}
\tsc{BEC}
\tsc{DE}

\begin{document}
\let\WriteBookmarks\relax

\shorttitle{EDARP}

\shortauthors{Sten Elling Tingstad Jacobsen et al.}

\title [mode = title]{Learning to Dial-a-Ride: A Deep Graph Reinforcement Learning Approach to the Electric Dial-a-Ride Problem}
                      

%
\author[1,2]{Sten Elling Tingstad Jacobsen}

\cormark[1]

\fnmark[1]

\ead{elling@chalmers.se}



\affiliation[1]{organization={Department of Electrical Engineering, Chalmers University of Technology},
    city={Gothenburg},
    country={Sweden}}

\author[1]{Attila Lischka}
\ead{lischka@chalmers.se}
\author[1]{Balázs Kulcsár}
\ead{kulcsar@chalmers.se}
\author[2]{Anders Lindman}
\ead{anders.lindman@volvocars.com}

\affiliation[2]{organization={Volvo Cars},
    city={Gothenburg},
    country={Sweden}}

\cortext[cor1]{Corresponding author}
\cortext[cor2]{Principal corresponding author}



\begin{abstract}
Urban mobility systems are transitioning toward electric, on-demand services, creating operational challenges for fleet management under energy and service-quality constraints. The Electric Dial-a-Ride Problem (E-DARP) extends the classical dial-a-ride problem by incorporating limited battery capacity and nonlinear charging dynamics, increasing computational complexity and limiting the scalability of exact methods for real-time use. This paper proposes a deep reinforcement learning approach based on an edge-centric graph neural network encoder and an attention-driven route construction policy. By operating directly on edge attributes such as travel time and energy consumption, the method captures non-Euclidean, asymmetric, and energy-dependent routing costs in real road networks. The learned policy jointly optimizes routing, charging, and service quality without relying on Euclidean assumptions or handcrafted heuristics. The approach is evaluated on two case studies using ride-sharing data from San Francisco. On benchmark instances, the method achieves solutions within 0.4\% of best-known results while reducing computation times by orders of magnitude. A second case study considers large-scale instances with up to 250 request pairs, realistic energy models, and nonlinear charging. On these instances, the learned policy outperforms Adaptive Large Neighborhood Search (ALNS) by 9.5\% in solution quality while achieving 100\% service completion, with inference times under 10 seconds compared to hours for the metaheuristic. Finally, sensitivity analyses quantify the impact of battery capacity, fleet size, ride-sharing capacity, and reward weights, while robustness experiments show that deterministically trained policies generalize effectively under stochastic conditions.
\end{abstract}




\maketitle
\section{Introduction}

Urban transportation systems are undergoing major changes as cities adopt electric vehicles, autonomous driving, and on-demand services. More than half of the world's population now lives in cities, and this share is expected to reach 68\% by 2050 \citep{un_urbanization_2018}. This rapid urbanization creates growing pressure on transportation networks to reduce emissions, ease congestion, and provide accessible mobility for all residents \citep{creutzig_transport_2015}. Transportation currently produces nearly a quarter of global energy-related CO$_2$ emissions \citep{iea_transport_2020}, making cleaner mobility solutions essential for climate action. Fleets of electric autonomous vehicles offer a way to address these challenges by combining zero-emission transportation with intelligent coordination for efficient, on-demand services \citep{fagnant_automated_2015, zhang_eamods_2016}. However, operating such systems requires solving complex optimization problems that coordinate vehicle routes, passenger assignments, and battery charging in real time, challenges that intensify as fleets grow larger and serve wider areas \citep{pavone_autonomous_2012, alonso-mora_demand_2017}.

The Electric Dial-a-Ride Problem (E-DARP), introduced by \citep{BONGIOVANNI2019436}, formalizes these challenges by extending the classical Dial-a-Ride Problem with electric vehicle constraints such as limited driving range and charging requirements. In this problem, a fleet of electric vehicles operates from one or more depots to serve passenger transportation requests, where each request consists of picking up passengers at an origin and delivering them to a destination. The problem incorporates user-centric service quality constraints: time windows bound when service can begin and end, maximum ride time limits constrain how long passengers spend in vehicles, and ordering constraints ensure pickups occur before deliveries. Additionally, vehicles have limited battery capacity and must visit charging stations when needed. These constraints, combined with vehicle capacity limits, create a computationally challenging problem. Even without electric constraints, the classical DARP is NP-hard, and a single vehicle serving $n$ requests admits $(2n)!/2^n$ possible solutions, approximately $2.38 \times 10^{18}$ for just $n=10$ requests \citep{cordeau2003tabu}.The integration of nonlinear charging dynamics and energy-dependent routing further compounds this complexity.

Despite this complexity, exact optimization methods for dial-a-ride problems have made significant progress. Branch-and-cut \citep{cordeau2006branch, braekers2014exact} and branch-and-price approaches \citep{ropke2009branch, gschwind2015effective} can now solve classical DARP instances with up to 8 vehicles and 96 requests. Recent work has extended these methods to electric vehicle settings, including branch-and-cut \citep{BONGIOVANNI2019436}, granular tabu search \citep{goeke2019granular}, robust optimization for demand uncertainty \citep{liu2023robust}, and branch-and-price with fragment-based representations \citep{su2024branch}. Despite these advances, the combinatorial complexity introduced by time windows, capacity limits, ride-time constraints, and nonlinear charging dynamics renders exact methods computationally prohibitive for large-scale, real-time applications requiring sub-second response times.
These scalability challenges have motivated growing interest in learning-based approaches. Deep reinforcement learning (DRL) has emerged as a promising paradigm for combinatorial optimization, offering an unsupervised learning approach that discovers effective policies through trial-and-error interaction rather than requiring labeled solutions from existing solvers \citep{bengio2021machine}. Foundational work on attention mechanisms \citep{vinyals2015pointer, bello2017neural} demonstrated effectiveness for vehicle routing \citep{nazari2018reinforcement, kool2019attention}. The POMO framework \citep{kwon2020pomo} showed that learned policies can match or exceed classical heuristics by exploiting multiple solution trajectories during training, an approach particularly relevant to E-DARP where sequential decision structure and robust policy learning align well with POMO's design. Extensions to pickup-and-delivery problems have introduced heterogeneous attention mechanisms respecting ordering constraints \citep{li2021heterogeneous} and symmetric architectures exploiting paired location structure \citep{luo2024solving}. However, these methods primarily address basic PDP or PDPTW without energy constraints, leaving the integration of battery management and charging decisions largely unexplored in the learning-based literature.

Despite advances in both exact methods and DRL, learning-based approaches have not been applied to the full E-DARP. Supervised learning requires high-quality solutions from existing solvers, limiting applicability to problems where such solvers exist \citep{vinyals2015pointer}. Most DRL research addresses either conventional DARP without energy constraints or electric vehicle routing without passenger-centric requirements such as ride-time limits \citep{schneider2014electric, keskin2016partial, basso2022dynamic}. This leaves a gap: no learning-based method currently handles the combination of battery management, charging decisions, and user-centric service constraints that defines the E-DARP.
At the same time, applying DRL to E-DARP presents distinct challenges. First, the combination of passenger time windows, vehicle capacity, ride-time limits, and battery constraints creates a complex constraint structure that standard neural architectures struggle to represent effectively. Second, and crucially, the E-DARP with energy-dependent routing is inherently non-Euclidean and asymmetric: traveling from location A to B differs from B to A due to factors such as elevation changes, one-way streets, and traffic patterns. These direction-dependent costs cannot be captured by coordinate-based distance calculations. Standard attention-based architectures for vehicle routing \citep{kool2019attention, kwon2020pomo} typically embed node locations as Euclidean coordinates and derive edge features (distances, travel times) from these coordinates. This approach fundamentally cannot represent asymmetric travel times, asymmetric real world energy consumption, or real-world road network structures which are asymmetric. Consequently, node-based graph neural networks and standard attention mechanisms are ill-suited for learning effective policies in energy-aware routing problems. Third, while DRL has shown promise on synthetic Euclidean benchmarks, scalability to operationally meaningful instance sizes with realistic, non-Euclidean cost structures remains underexplored. Most existing work evaluates on instances with fewer than 100 requests, leaving a gap between academic benchmarks and the larger problems encountered in practice. Finally, the interpretability and transparency of learned policies presents an ongoing challenge. Neural network-based approaches are often criticized as black-box models, making it difficult to understand why specific routing or charging decisions are made \citep{rudin2019stop, molnar2020interpretable}. In safety-critical transportation applications, this opacity can hinder trust, regulatory acceptance, and debugging of failure cases \citep{arrieta2020explainable}. While attention mechanisms offer some degree of interpretability by revealing which problem elements influence decisions \citep{vaswani2017attention, kool2019attention}, systematic frameworks for explaining DRL policies in vehicle routing remain underdeveloped \citep{glanois2024survey}. Recent work has made progress through post-hoc explanation methods \citep{puiutta2020explainable}, mechanistic interpretability of neural routing solvers \citep{narad2025mechanistic}, and interpretable policy architectures \citep{silva2020optimization}, suggesting that learning-based methods are becoming increasingly amenable to interpretation, though this remains an active area of research beyond the scope of the present work.

This work applies the GREAT (Graph Edge Attention Network) encoder architecture \citep{lischka2025greatarchitectureedgebasedgraph} to the E-DARP within an unsupervised deep reinforcement learning framework. Unlike supervised approaches that require optimal or near-optimal solutions for training, our method learns routing policies entirely through reward-driven exploration, eliminating dependence on existing solvers. GREAT is specifically designed for non-Euclidean routing problems. Conventional node-based architectures derive edge information from node coordinates, but GREAT operates directly on edges: edge features serve as the primary inputs to the attention mechanism, and information propagates between edges that share endpoints. This design naturally represents asymmetric costs and direction-dependent energy consumption. For E-DARP, routing decisions correspond to selecting edges between pickup locations, delivery locations, and charging stations, making the edge-based formulation a natural fit.

The main contributions of this study are threefold:
\begin{enumerate}
    \item \textbf{Unsupervised learning framework with edge-based graph neural architecture.} We demonstrate that the GREAT encoder architecture, combined with deep reinforcement learning, enables fully unsupervised learning of E-DARP routing policies. The model discovers effective strategies through reward-driven exploration without requiring labeled solutions from optimization algorithms. By processing routing problems through edge-level attention rather than node-based representations and operating directly on edge features (travel times, distances, energy consumption), the architecture naturally handles asymmetric costs and non-Euclidean network structures that characterize real-world electric vehicle routing. This design choice, combined with an attention-based decoder and POMO training, enables the policy to learn sophisticated routing and charging strategies.
    
    \item \textbf{Computational efficiency and optimality.} We evaluate on two complementary case studies. On benchmark instances from \citep{BONGIOVANNI2019436}, our method achieves optimal solutions (0.0\% gap) for problems with up to 30 requests and maintains near-optimal quality (0.4\% gap) for 50-request instances, while providing computational speedups of 20$\times$ to over 7,000$\times$ compared to exact methods. On a second case study with realistic energy models and nonlinear charging across greater San Francisco, curriculum learning enables scaling to 250 request pairs (500 nodes). On these instances, the learned policy outperforms Adaptive Large Neighborhood Search by 9.5\% in solution quality while achieving 100\% service completion, with inference times under 10 seconds compared to hours for the metaheuristic.
    
    \item \textbf{Comprehensive sensitivity analysis.} We conduct systematic analyses of operational factors (battery capacity, fleet size, ride-sharing capacity), reward function weights, neural architecture hyperparameters, and robustness under stochastic conditions. Key findings include that larger batteries improve profits by 6 to 15\%, adequate completion incentives are essential for full service coverage, smaller architectures match or outperform larger models, and deterministically trained policies generalize effectively to 10\% operational uncertainty.
\end{enumerate}
Together, these contributions demonstrate that learning-based, edge-centric models can tackle large-scale, energy-aware routing problems efficiently and adaptively, providing a practical framework for real-world autonomous fleet deployment.
The remainder of this paper is organized as follows. We begin with an introductory example illustrating the problem's computational challenges, then formulate the E-DARP as a Markov Decision Process and provide a detailed mathematical problem statement including constraints and objectives (Section \ref{sec:problem}). The methodology is then presented, describing the GREAT encoder architecture, the attention-based decoder, feasibility masking, and the reward function design (Section \ref{sec:methodology}). Next, we detail the experimental setup, including problem instance generation, training procedures, and baseline methods for both benchmark and large-scale case studies (Section \ref{sec:experimental_setup}). Comprehensive results and analysis follow, covering benchmark performance comparison against exact methods, large-scale curriculum learning evaluation, sensitivity analysis across battery sizes, fleet configurations, and ride-sharing capacities, hyperparameter analysis, reward weight sensitivity, and robustness evaluation under stochastic conditions (Section \ref{sec:results}). Finally, we conclude with a summary of findings, discussion of limitations, and directions for future research (Section \ref{sec:conclusion}).

\section{Electric Dial-a-Ride Problem: Definition and Formulation}
\label{sec:problem}

This section formulates the Electric Dial-a-Ride Problem for deep reinforcement learning. We present a motivating example illustrating the problem's computational challenges (Section \ref{sec:example}), formulate the sequential decision-making process as a Markov Decision Process (Section \ref{sec:mdp}), and provide a detailed mathematical specification of constraints and objectives (Section \ref{sec:problem_statement}). 

\subsection{Introductory Example}
\label{sec:example}

\begin{figure}
    \centering
    \includegraphics[width=0.7\textwidth]{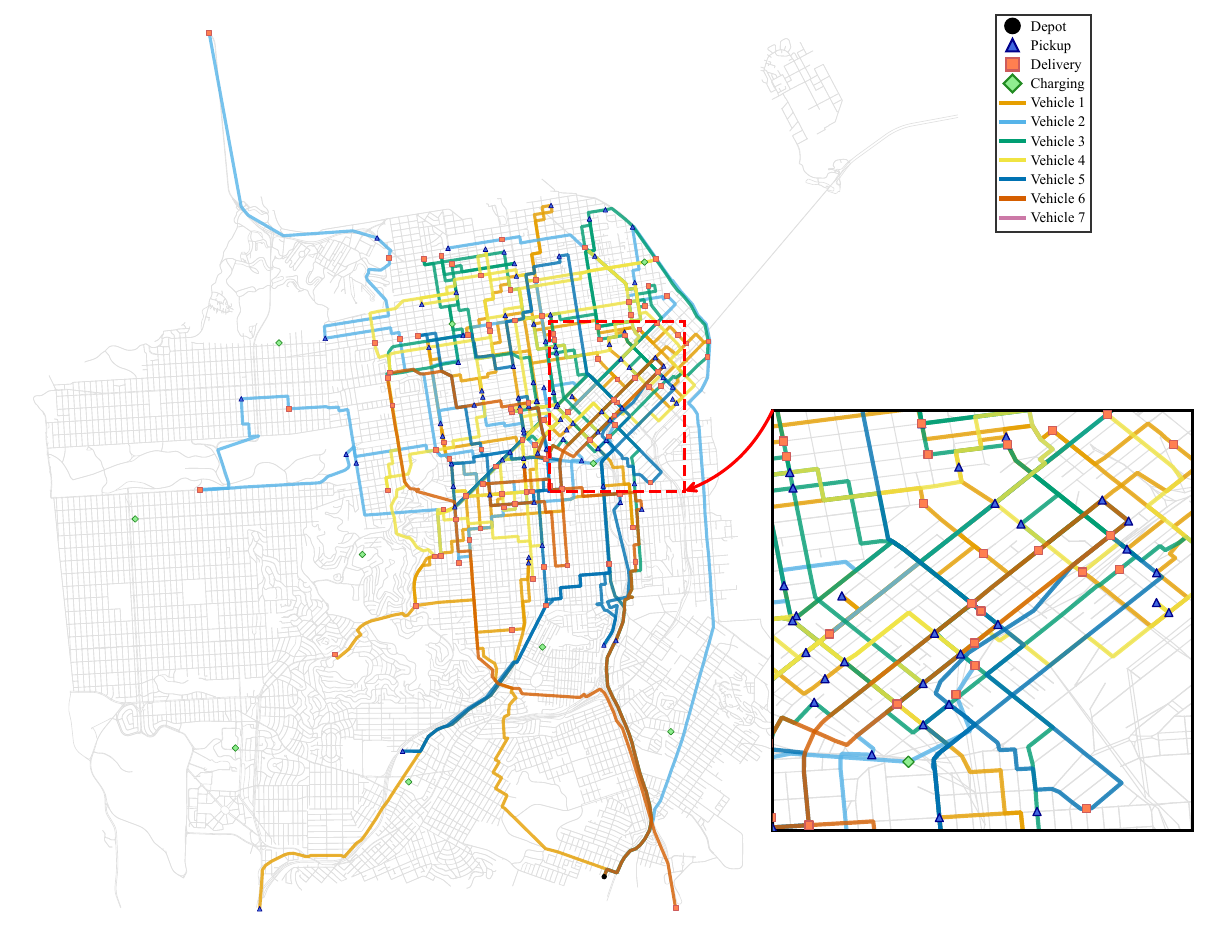}
    \caption{Example E-DARP solution with seven vehicles serving pickup--delivery requests and visiting charging stations in San Francisco.}
    \label{fig:edarp_example}
\end{figure}

To illustrate the computational challenges of the Electric Dial-a-Ride Problem (E-DARP), consider a representative instance with seven electric vehicles serving 100 customer requests over a 12-hour planning horizon. Each vehicle has a 40 kWh battery, a maximum capacity of three passengers, and access to ten charging stations distributed across the service area. The objective is to construct feasible vehicle routes that serve all requests while minimizing energy consumption, waiting time, and passenger ride time. Fig. \ref{fig:edarp_example} illustrates a solution to such an instance.

Even this seemingly moderate scenario presents fundamental challenges that make the problem computationally difficult. The first challenge is combinatorial explosion: the number of feasible solutions grows exponentially with the number of requests. Even under highly simplified assumptions where each of the 100 requests can be assigned to one of seven vehicles or left unserved, there are at least $8^{100} \approx 1.3 \times 10^{90}$ possible assignment combinations. For any fixed assignment, the number of feasible pickup-delivery sequences grows factorially; even for a single vehicle serving $n$ requests, there are $(2n)! / 2^n$ precedence-feasible orderings \citep{cordeau2003tabu}. Accounting for multiple vehicles, time windows, capacity limits, battery constraints, and charging decisions imposes further restrictions this solution space \citep{BONGIOVANNI2019436}. The second challenge is the tight coupling between routing and charging decisions \citep{schneider2014electric, keskin2016partial}. A vehicle may choose to charge early, incurring idle time that delays service, or serve nearby requests first, risking battery depletion and infeasibility later in the route. Greedy strategies that select the nearest feasible request can strand vehicles in low-demand areas with insufficient battery to reach charging stations or complete additional requests \citep{sweda2017adaptive}. The third challenge is real-time requirements: dynamic fleet operations must repeatedly recompute decisions as new requests arrive and conditions change \citep{pillac2013review}. Exact mixed-integer optimization methods often require minutes to hours for instances of this size \citep{su2024branch, gschwind2015effective}, making them unsuitable for real-time deployment.

These challenges motivate our approach of formulating the E-DARP as a Markov Decision Process, where an agent learns to construct vehicle routes through sequential decisions that balance service quality, battery management, and operational efficiency without exhaustive enumeration of the solution space.

\subsection{Markov Decision Process Formulation}
\label{sec:mdp}
The E-DARP can be represented as a Markov Decision Process (MDP) denoted by $\mathcal{M} = (\mathcal{S}, \mathcal{A}, \mathcal{R}, \mathcal{T}, H)$. This formulation encompasses the state space $\mathcal{S}$ with states $s_t \in \mathcal{S}$, the action space $\mathcal{A}$ with actions $a_t \in \mathcal{A}$, the reward function $\mathcal{R}(\cdot)$, the transition function $\mathcal{T}(\cdot)$, and the horizon $H$ indicating the episode duration. The state of a problem instance encodes the current vehicle's operational status (location, time, battery level, passenger load) and the set of served requests, where the next state $s_{t+1}$ is determined by applying the chosen action $a_t$ (selecting the next node to visit) to the current state $s_t$. An agent is introduced in the MDP to construct solutions by learning a policy $\pi_\theta(a|s)$. The standard objective in reinforcement learning seeks a policy that generates a trajectory $\xi$ maximizing the expected reward: $\pi^* = \arg\max_{\pi_\theta} \mathbb{E}_{\xi \sim \pi_\theta}[\mathcal{R}(\xi)]$. Let $\mathcal{C}(\xi)$ denote the total cost and $\mathcal{P}(\xi)$ the total profit associated with trajectory $\xi$; the reward is then defined as $\mathcal{R}(\xi) = \mathcal{P}(\xi) - \mathcal{C}(\xi)$, assigned at episode termination. However, real-world applications typically allow generating multiple candidate solutions rather than committing to a single trajectory. For combinatorial optimization problems, the learning objective should therefore maximize the best solution among $K$ attempts: $\pi^* = \arg\max_{\pi_\theta} \mathbb{E}_{\xi_i \sim \pi_\theta}[\max_{i=1,\ldots,K} \mathcal{R}(\xi_i)]$. This multi-shot formulation motivates learning diverse policies that maintain sufficient exploration across solution attempts, allowing the model to discover high-quality solutions with high probability rather than converging prematurely to a single deterministic trajectory.
\subsection{Combinatorial Optimization Formulation}
\label{sec:problem_statement}

We formulate the E-DARP as a combinatorial optimization problem that determines vehicle routes, service times, and charging decisions to minimize operational costs while satisfying passenger service requirements and electric vehicle constraints. An E-DARP instance with $n$ transportation requests is defined on a directed graph $G=(V,E)$ where $V = \{0\} \cup P \cup D \cup F$ comprises the depot (node 0), pickup nodes $P=\{1,\dots,n\}$, delivery nodes $D=\{n+1,\dots,2n\}$, and charging stations $F$. Each request $i \in \{1,\dots,n\}$ pairs pickup node $i$ with delivery node $n+i$. Edges $(i,j) \in E$ are characterized by travel time $\delta_{ij}$, distance $d_{ij}$, and energy consumption $\varepsilon_{ij}$. Nodes $i \in P \cup D \cup F$ have service time $\sigma_i$ and time window $[a_i, \ell_i]$. Pickup nodes have passenger load $q_i > 0$ (removed at corresponding delivery), and requests impose maximum ride time $\bar{L}_i$. A fleet of $K$ homogeneous electric vehicles operates from the depot, each with passenger capacity $Q$ and battery capacity $B$. Let $\rho^k = (v^k_0, v^k_1, \ldots, v^k_{T_k})$ denote the route for vehicle $k$, where $v^k_t \in V$ is the node visited at position $t$ with $v^k_0 = v^k_{T_k} = 0$. Let $t^k_t \in \mathbb{R}_+$ denote the service start time and $b^k_t \in [0,1]$ the state of charge after visiting node $v^k_t$. The variable $l^k_t$ denotes the current onboard passenger load of vehicle $k$ at step $t$, while $q_i$ represents the load change incurred when visiting node $i$ (positive for pickups 
and negative for deliveries). The objective function combines energy consumption, passenger waiting time at pickups, and delivery lateness:
\begin{align}
J_e(\{\rho^k\}) &= \sum_{k=1}^K \sum_{t=0}^{T_k-1} \varepsilon_{v^k_t v^k_{t+1}}, \label{eq:energy_cost} \\
J_w(\{\rho^k\}) &= \sum_{k=1}^K \sum_{t: v^k_t \in P} \max(0, a_{v^k_t} - t^k_{t-1} - \delta_{v^k_{t-1} v^k_t}), \label{eq:waiting_cost} \\
J_l(\{\rho^k\}) &= \sum_{k=1}^K \sum_{t: v^k_t \in D} \max(0, t^k_t - \ell_{v^k_t}), \label{eq:lateness_cost}
\end{align}
yielding the total cost $J(\{\rho^k\}) = w_e J_e + w_w J_w + w_l J_l$, where $w_e, w_w, w_l \geq 0$ are weight parameters. The optimization problem is:

\begin{subequations}
\label{eq:edarp_optimization}
\begin{alignat}{2}
\minimize_{\{\rho^k\}_{k=1}^K, \{t^k_t\}, \{b^k_t\}} \quad
& J(\{\rho^k\})
\label{eq:edarp_obj} \\[0.5em]
\text{s.t.} \quad
& n_{\text{served}}(\{\rho^k\}) = n,
\label{eq:edarp_service} \\
& i, n+i \in \bigcup_{k=1}^K \rho^k, \quad i \prec n+i
&& \forall i \in \{1,\ldots,n\},
\label{eq:edarp_precedence} \\
& 0 \leq \sum_{s=1}^{t} q_{v^k_s} \leq Q
&& \forall k,\ \forall t,
\label{eq:edarp_capacity} \\
& t^k_{t+1}
= \max(t^k_t + \delta_{v^k_t v^k_{t+1}}, a_{v^k_{t+1}})
+ \sigma_{v^k_{t+1}}
&& \forall k,\ \forall t,
\label{eq:edarp_timeprop} \\
& a_{v^k_t} \leq t^k_t \leq \ell_{v^k_t}
&& \forall k,\ \forall t,
\label{eq:edarp_timewindow} \\
& t^k_{t'} - t^k_t \leq \bar{L}_i
&& \forall i,\ k:\ v^k_t=i,\ v^k_{t'}=n+i,
\label{eq:edarp_ridetime} \\
& b^k_{t+1}
= b^k_t - \varepsilon_{v^k_t v^k_{t+1}}/B + \Delta^k_{t+1}
&& \forall k,\ \forall t,
\label{eq:edarp_battery} \\
& 0 \leq b^k_t \leq 1
&& \forall k,\ \forall t,
\label{eq:edarp_soc} \\
& b^k_0 = 1
&& \forall k,
\label{eq:edarp_socinit} \\
& \Delta^k_t =
\begin{cases}
P^c(b^k_{t-1}) \, \sigma_{v^k_t} / B, & v^k_t \in F, \\
0, & \text{otherwise},
\end{cases}
&& \forall k,\ \forall t.
\label{eq:edarp_charging}
\end{alignat}
\end{subequations}

where the charging power function is:

\begin{equation}
P^c(b_t) = \begin{cases}
100 \text{ kW} & \text{if $b_t$} < 0.45 \\
100 - 140 \cdot (\text{$b_t$} - 0.45) \text{ kW} & \text{if } 0.45 \leq \text{$b_t$} \leq 0.95 \\
30 \text{ kW} & \text{if $b_t$} > 0.95
\end{cases}
\label{eq:charging_power}
\end{equation}

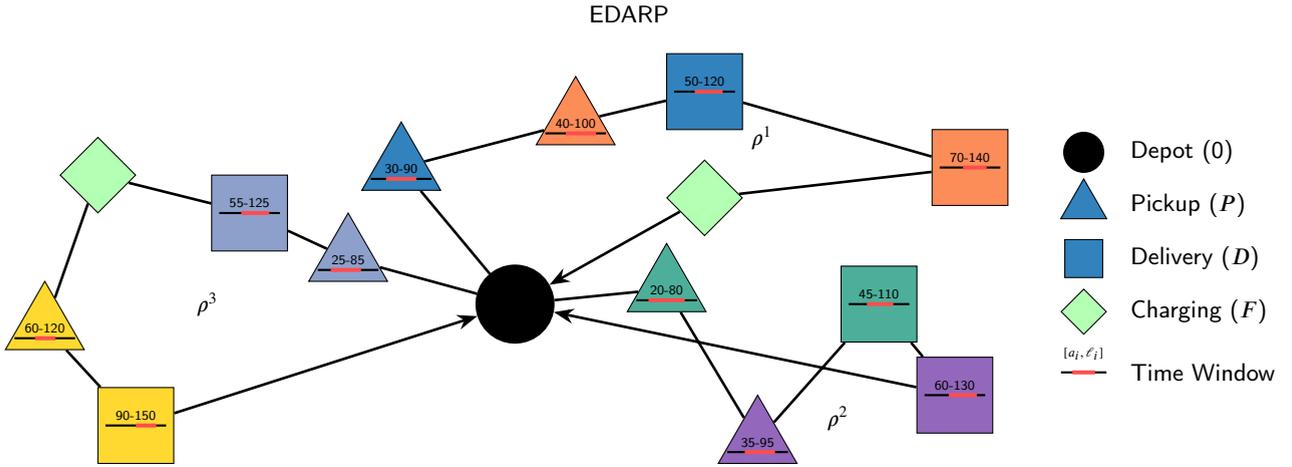
\begin{figure}
  \centering
\begin{tikzpicture}[
    >=Stealth,
    font=\small,
    route/.style={->, line width=1pt, draw=black},
    node size/.style={minimum size=10mm, inner sep=0pt},
    dep/.style={circle, draw, very thick, fill=white, node size},
    tri/.style={regular polygon, regular polygon sides=3, draw, minimum size=12mm, inner sep=0pt},
    sq/.style={rectangle, draw, minimum size=10mm, inner sep=0pt},
    charge/.style={diamond, draw, node size, fill=green!30},
    tw outer/.style={draw=black, line width=0.8pt},
    tw inner/.style={draw=red!70, line width=1.5pt},
    tw label/.style={font=\fontsize{4}{5}\selectfont, inner sep=0.5pt},
    id label/.style={font=\tiny\bfseries, fill=white, inner sep=1pt, rounded corners=1pt},
]
\definecolor{req1}{RGB}{49,130,189}
\definecolor{req2}{RGB}{252,141,89}
\definecolor{req3}{RGB}{77,175,153}
\definecolor{req4}{RGB}{146,103,188}
\definecolor{req5}{RGB}{141,160,203}
\definecolor{req6}{RGB}{255,217,47}

\node (depot) [dep, fill=black] at (0,0) {};

\node (c1) [charge] at (2.5,1.4) {};
\node (c2) [charge] at (-5.5,1.7) {};

\node (p1) [tri, fill=req1] at (-1.5,1.8) {};
\node (d1) [sq,  fill=req1] at (2.5,2.8) {};
\node (p2) [tri, fill=req2] at (0.8,2.4) {};
\node (d2) [sq,  fill=req2] at (6.0,1.8) {};
\draw[route] (depot) -- (p1) -- (p2) -- (d1)  -- (d2) -- (c1) -- (depot);

\node (p3) [tri, fill=req3] at (2.0,0.2) {};
\node (d3) [sq,  fill=req3] at (4.8,-0.0) {};
\node (p4) [tri, fill=req4] at (3.2,-1.8) {};
\node (d4) [sq,  fill=req4] at (5.8,-1.2) {};
\draw[route] (depot) -- (p3) -- (p4) -- (d3) -- (d4) -- (depot);

\node (p5) [tri, fill=req5] at (-2.2,0.6) {};
\node (d5) [sq,  fill=req5] at (-3.5,1.2) {};
\node (p6) [tri, fill=req6] at (-6.2,-0.3) {};
\node (d6) [sq,  fill=req6] at (-5.0,-1.6) {};
\draw[route] (depot) -- (p5) -- (d5) -- (c2) -- (p6) -- (d6) -- (depot);


\begin{scope}[shift={(p1)}]
  \draw[tw outer] (-0.4,-0.15) -- ++(0.8,0);
  \draw[tw inner] (-0.4,-0.15) ++(0.2,0) -- ++(0.4,0);
  \node[tw label, above] at (0,-0.1) {30-90};
\end{scope}

\begin{scope}[shift={(d1)}]
  \draw[tw outer] (-0.4,0) -- ++(0.8,0);
  \draw[tw inner] (-0.4,0) ++(0.27,0) -- ++(0.37,0);
  \node[tw label, above] at (0,0.05) {50-120};
\end{scope}

\begin{scope}[shift={(p2)}]
  \draw[tw outer] (-0.4,-0.15) -- ++(0.8,0);
  \draw[tw inner] (-0.4,-0.15) ++(0.27,0) -- ++(0.4,0);
  \node[tw label, above] at (0,-0.1) {40-100};
\end{scope}

\begin{scope}[shift={(d2)}]
  \draw[tw outer] (-0.4,0) -- ++(0.8,0);
  \draw[tw inner] (-0.4,0) ++(0.31,0) -- ++(0.31,0);
  \node[tw label, above] at (0,0.05) {70-140};
\end{scope}

\begin{scope}[shift={(p3)}]
  \draw[tw outer] (-0.4,-0.15) -- ++(0.8,0);
  \draw[tw inner] (-0.4,-0.15) ++(0.16,0) -- ++(0.48,0);
  \node[tw label, above] at (0,-0.1) {20-80};
\end{scope}

\begin{scope}[shift={(d3)}]
  \draw[tw outer] (-0.4,0) -- ++(0.8,0);
  \draw[tw inner] (-0.4,0) ++(0.24,0) -- ++(0.35,0);
  \node[tw label, above] at (0,0.05) {45-110};
\end{scope}

\begin{scope}[shift={(p4)}]
  \draw[tw outer] (-0.4,-0.15) -- ++(0.8,0);
  \draw[tw inner] (-0.4,-0.15) ++(0.23,0) -- ++(0.4,0);
  \node[tw label, above] at (0,-0.1) {35-95};
\end{scope}

\begin{scope}[shift={(d4)}]
  \draw[tw outer] (-0.4,0) -- ++(0.8,0);
  \draw[tw inner] (-0.4,0) ++(0.32,0) -- ++(0.37,0);
  \node[tw label, above] at (0,0.05) {60-130};
\end{scope}

\begin{scope}[shift={(p5)}]
  \draw[tw outer] (-0.4,-0.15) -- ++(0.8,0);
  \draw[tw inner] (-0.4,-0.15) ++(0.17,0) -- ++(0.4,0);
  \node[tw label, above] at (0,-0.1) {25-85};
\end{scope}

\begin{scope}[shift={(d5)}]
  \draw[tw outer] (-0.4,0) -- ++(0.8,0);
  \draw[tw inner] (-0.4,0) ++(0.29,0) -- ++(0.37,0);
  \node[tw label, above] at (0,0.05) {55-125};
\end{scope}

\begin{scope}[shift={(p6)}]
  \draw[tw outer] (-0.4,-0.15) -- ++(0.8,0);
  \draw[tw inner] (-0.4,-0.15) ++(0.27,0) -- ++(0.27,0);
  \node[tw label, above] at (0,-0.1) {60-120};
\end{scope}

\begin{scope}[shift={(d6)}]
  \draw[tw outer] (-0.4,0) -- ++(0.8,0);
  \draw[tw inner] (-0.4,0) ++(0.4,0) -- ++(0.27,0);
  \node[tw label, above] at (0,0.05) {90-150};
\end{scope}

\begin{scope}[shift={(7.5,2.0)}]
  \node[dep, minimum size=5mm, fill=black] (Ldep) at (0,0) {};
  \node[anchor=west, font=\small] at (0.5,0) {Depot (0)};
  
  \node[tri, fill=req1, minimum size=7mm] (Ltri) at (0,-0.7) {};
  \node[anchor=west, font=\small] at (0.5,-0.7) {Pickup ($P$)};
  
  \node[sq,  fill=req1, minimum size=5mm] (Lsq) at (0,-1.4) {};
  \node[anchor=west, font=\small] at (0.5,-1.4) {Delivery ($D$)};
  
  \node[charge, minimum size=6mm] (Lcharge) at (0,-2.1) {};
  \node[anchor=west, font=\small] at (0.5,-2.1) {Charging ($F$)};
  
  \draw[tw outer] (-0.3,-2.9) -- ++(0.6,0);
  \draw[tw inner] (-0.3,-2.9) ++(0.15,0) -- ++(0.3,0);
  \node[font=\fontsize{5}{6}\selectfont, above] at (0,-2.85) {$[a_i,\ell_i]$};
  \node[anchor=west, font=\small] at (0.5,-2.9) {Time Window};
\end{scope}

\node[anchor=west]  at (3.0,2.2)  {$\rho^1$};
\node[anchor=west]  at (4.0,-1.5) {$\rho^2$};
\node[anchor=east]  at (-3.8,0.0) {$\rho^3$};
\end{tikzpicture}
  \caption{Example E-DARP solution with three vehicle routes and time windows. The depot (node 0) is shown as a filled black circle. Each request $i$ consists of a pickup node $i \in P$ (triangle) and delivery node $n+i \in D$ (square) in matching colors. Black horizontal bars represent the planning horizon for each node, while the overlaid red bars indicate the feasible service time windows $[a_i, \ell_i]$ during which the vehicle must arrive. Green diamonds represent charging stations $f \in F$ that are available throughout the planning horizon. Routes $\rho^1$ and $\rho^3$ include charging stops to maintain battery feasibility throughout their journeys. Each vehicle can carry up to $Q$ passengers simultaneously while respecting both time window constraints and passenger ride time limits.}
  \label{fig:edarptw-routes}
\end{figure}

Constraint \eqref{eq:edarp_service} enforces complete service coverage by requiring that all $n$ requests are served. 
Constraint \eqref{eq:edarp_precedence} ensures that each request is served with its pickup preceding the corresponding delivery. Vehicle capacity limits are imposed by Constraint \eqref{eq:edarp_capacity}, which bounds the onboard passenger load at all times.Temporal feasibility is enforced through Constraints \eqref{eq:edarp_timeprop} and \eqref{eq:edarp_timewindow}, which define service time propagation along routes and ensure adherence to node time windows. Passenger service quality is controlled by Constraint \eqref{eq:edarp_ridetime}, which limits the maximum ride time for each request. Battery dynamics are governed by Constraints \eqref{eq:edarp_battery}--\eqref{eq:edarp_charging}, modeling energy consumption during travel, charging at stations, initial battery conditions, and state-of-charge bounds. For the reinforcement learning formulation, we relax the hard service requirement to accommodate resource-constrained scenarios where battery capacity or fleet size may be insufficient to serve all requests. The reward function becomes:

\begin{equation}
\mathcal{R}(\{\rho^k\}) = -J(\{\rho^k\}) + w_c \cdot n_{\text{served}}(\{\rho^k\})
\label{eq:reward}
\end{equation}

where $J(\{\rho^k\})$ is the objective value in Eq. \eqref{eq:edarp_optimization} and $w_c \geq 0$ weights the completion bonus. This formulation converts the hard constraint $n_{\text{served}}(\{\rho^k\}) = n$ into a soft objective that incentivizes maximum service coverage while allowing partial solutions when constraints prevent serving all demand. The agent receives $\mathcal{R}(\{\rho^k\})$ only at episode termination. Fig. \ref{fig:edarptw-routes} illustrates an example solution, and Table \ref{tab:edarptw_notation} summarizes notation.

\begin{table}[t]
\caption{Summary of notation and symbols used in the Electric Dial-a-Ride Problem (E-DARP) formulation.}
\label{tab:edarptw_notation}
\centering
\begin{tabular}{ll}
\hline
\textbf{Symbol} & \textbf{Description} \\ \hline
\multicolumn{2}{l}{\textit{Problem Structure}} \\
$n$ & Number of requests \\
$0$ & Depot node \\
$P=\{1,\dots,n\}$ & Pickup nodes \\
$D=\{n+1,\dots,2n\}$ & Delivery nodes \\
$F$ & Charging stations \\
$V=\{0\}\cup P\cup D\cup F$ & All nodes \\
$E \subseteq V \times V$ & All edges \\
$K$ & Number of vehicles \\ \hline
\multicolumn{2}{l}{\textit{Vehicle State}} \\
$l^k_t$ & Number of passengers on board at step $t$ \\
$b_t \in [0,1]$ & State of Charge (SoC) at step $t$ \\
$\tau_t$ & Current time at step $t$ \\
$v_t$ & Current node at step $t$ \\ \hline
\multicolumn{2}{l}{\textit{Node Properties}} \\
$q_i$ & Passenger load change at node $i$ \\
$\bar{L}_i$ & Maximum ride time for request $i$ \\
$[a_i, \ell_i]$ & Time window at node $i$ \\
$\sigma_i$ & Service time at node $i$ \\ \hline
\multicolumn{2}{l}{\textit{Edge Properties}} \\
$\delta_{ij}$ & Travel time on edge $(i,j)$ \\
$d_{ij}$ & Distance on edge $(i,j)$ \\
$\varepsilon_{ij}$ & Energy consumption on edge $(i,j)$ \\ \hline
\multicolumn{2}{l}{\textit{Vehicle Parameters}} \\
$Q$ & Passenger capacity \\
$B$ & Battery capacity \\
$P^\text{c}$ & Charging power function \\ \hline
\end{tabular}
\end{table}
\section{Methodology}
\label{sec:methodology}

We solve the E-DARP using deep reinforcement learning with a neural network policy that constructs routes iteratively. The neural network encoder processes the problem instance structure (Section \ref{sec:encoder}), while an attention-based decoder selects nodes to visit (Section \ref{sec:decoder}). We enforce feasibility through action masking rules (Section \ref{sec:mask}) and define the training objective based on solution cost (Section \ref{sec:reward}). 

\subsection{Problem Encoder}
\label{sec:encoder}
We use the GREAT (Graph Edge Attention Network) encoder \citep{lischka2025greatarchitectureedgebasedgraph} to process the problem instance into node embeddings. GREAT is a graph neural network that performs edge-level message passing rather than traditional node-level operations. This design is particularly well-suited for routing problems since such problems are typically defined by graph edge attributes. For example, such characterizing edge features are distance, travel time or expected energy consumption when traveling between a pair of nodes.

In particular, for a given E-DARP instance, we build an edge-based graph the following way:
For each node $i \in V$, we construct a feature vector that encodes the node type (depot, pickup, delivery, or charging station), the spatial coordinates normalized to the unit square, the time window $[a_i, \ell_i]$ normalized by the planning horizon, the service time $\sigma_i$, and the load change $q_i$ for pickups and deliveries. For each edge $(i,j) \in E$, we construct a feature vector containing the travel time $\delta_{ij}$, the distance $d_{ij}$, and the energy consumption $\varepsilon_{ij}$ expressed as a fraction of the battery capacity $B$.
Since GREAT is only edge-based and does not operate on node features, we translate all node features into additional edge attributes as outlined by \citep{lischka2025greatarchitectureedgebasedgraph}. 
This is done by appending the node feature of a node $i \in V$ to all edge features $\mathbf{h}_{ji}$ of edges $(j,i) \in E$ that lead to node $i$. Intuitively, this makes sense since, e.g., a load change $q_i$ for visiting node $i$ also corresponds to a load change $q_i$ when traversing the graph from node $j$ to $i$, i.e. taking edge $(j,i)$. 
Consequently, in total, each edge feature $\mathbf{h}_{ij}$ captures information about edge $(i,j) \in E$ and node $j \in V$. 
As a result, we obtain a graph where all relevant information is encoded in the graph's edges.
We note that all spatial features such as coordinates, distances, and times are normalized to the range $[0,1]$ using instance-specific min-max scaling. This ensures that the attention weights remain consistent across problem instances of different scales and ensures stable learning.

The GREAT encoder consists of several stacked individual GREAT layers (in particular, we use the GREAT NB layer version of \citep{lischka2025greatarchitectureedgebasedgraph}). Each layer consists of an edge-based multi-head attention sublayer, a feed forward layer, residual layers and normalizations. We refer the reader to \citep{lischka2025greatarchitectureedgebasedgraph} for details.
Here it suffices to say that at each layer $\ell$, the representation of edge$(i,j)$ is updated by attending to neighboring edges that share an endpoint with it. Specifically, we compute
\[
\tilde{\mathbf{h}}^{(\ell+1)}_{ij} = \operatorname{LN}\left(\mathcal{E}_A\left(\mathbf{h}^{(\ell)}_{ij}, \{\mathbf{h}^{(\ell)}_{km} : k = i \text{ or } m = j\}\right) + \mathbf{h}^{(\ell)}_{ij}\right),
\]
where $\mathbf{h}^{(\ell)}_{ij}$ denotes the feature vector for edge$(i,j)$ at layer $\ell$, and the edge attention mechanism $\mathcal{E}_A$ aggregates information from all edges that share either the source node $i$ or the target node $j$. The $\mathcal{LN}(\cdot)$ denote the layer normalization operator.  This edge-focused information passing allows the model to learn which routing decisions are promising by directly comparing adjacent edges in the graph structure. 
Afterwards, the final representation of layer $\ell$ is computed as
\[
\mathbf{h}^{(\ell+1)}_{ij} = \mathcal{LN}\left((\mathbf{W}'\sigma(\mathbf{W}''\tilde{\mathbf{h}}^{(\ell+1)}_{ij} + \mathbf{b}'') + \mathbf{b}') + \tilde{\mathbf{h}}^{(\ell+1)}_{ij}\right),
\]
where $\mathbf{W}', \mathbf{W}'', \mathbf{b}', \mathbf{b}''$ are trainable weights and $\sigma$ is the ReLU activation function.
After processing through all layers, we derive node embeddings by aggregating the representations of all incoming edges:
\[
\mathbf{z}_i = \sum_{j \in V} \omega_{ji} \mathbf{h}^{(L)}_{ji}
\]
where $L$ is the number of attention layers and $\omega_{ji}$ is a learned attention score that reflects the significance of each edge feature $\mathbf{h}^{(L)}_{ji}$ for node $i$. This produces a matrix $\mathbf{Z} \in \mathbb{R}^{N \times d_h}$ where $N$ is the total number of nodes and $d_h$ is the embedding dimension. These node embeddings are afterwards passed on to the route decoder. We note that the edge embeddings produced by the GREAT encoder are subsequently aggregated into node embeddings. This transformation is necessary to reduce both memory requirements and the size of the action space during the decoding procedure. In particular, it reduces the representation from $\mathcal{O}(n^2)$ edge embeddings to $\mathcal{O}(n)$ node embeddings, enabling efficient autoregressive decoding over candidate next nodes. We emphasize that this aggregation is performed only after edge-level reasoning has been completed. Directly operating on node features from the outset is ill-suited for energy-aware routing problems, as it cannot faithfully represent asymmetric, direction-dependent costs such as energy consumption that varies with travel direction due to terrain gradients or traffic patterns.


\subsection{Route Decoder}
\label{sec:decoder}

The decoder constructs routes step by step using an attention-based pointer mechanism. At each decision step $t$, the decoder produces a probability distribution over the next node to visit, conditioned on the current partial route and the operational state of the vehicle.

We first build a context-aware query vector $\mathbf{c}_t \in \mathbb{R}^{d_h}$ that captures the current situation. The query combines information from multiple sources through learned linear projections. We add the embedding of the current node weighted by $\mathbf{W}_{\text{curr}}$, the depot embedding weighted by $\mathbf{W}_{\text{depot}}$, the average of all node embeddings weighted by $\mathbf{W}_{\text{graph}}$, the average of visited node embeddings weighted by $\mathbf{W}_{\text{visited}}$, a summary of visited nodes based on the mask weighted by $\mathbf{W}_{\text{mask}}$, and projections of the operational state variables. The operational state includes the current passenger load $l_t$, battery level $b_t$, and time $\tau_t$, each projected through learned weight matrices $\mathbf{W}_l$, $\mathbf{W}_b$, and $\mathbf{W}_t$ respectively. Formally, we compute
\begin{align}
\mathbf{c}_t = &\; \mathbf{W}_{\text{curr}} \mathbf{z}_{v_t} + \mathbf{W}_{\text{depot}} \mathbf{z}_0 + \mathbf{W}_{\text{graph}} \bar{\mathbf{z}} + \mathbf{W}_{\text{visited}} \bar{\mathbf{z}}_{\text{visited}} \nonumber\\
      &+ \mathbf{W}_{\text{mask}} \phi(M_t, \mathbf{Z}) + \mathbf{W}_l l_t + \mathbf{W}_b b_t + \mathbf{W}_t \tau_t,
\end{align}
where $\mathbf{z}_{v_t}$ is the embedding of the current node, $\mathbf{z}_0$ is the depot embedding, $\bar{\mathbf{z}} = \frac{1}{N}\sum_{i} \mathbf{z}_i$ is the mean of all node embeddings, $\bar{\mathbf{z}}_{\text{visited}} = \frac{1}{|\mathcal{V}_t|}\sum_{i \in \mathcal{V}_t} \mathbf{z}_i$ is the mean of visited node embeddings, and $\phi(M_t, \mathbf{Z})$ summarizes which nodes have been masked based on the current feasibility constraints.

For each candidate next node $j$, we compute an attention score that balances semantic compatibility with energy efficiency. The score consists of two terms. The first term measures how well node $j$ fits the current context through the inner product of the query and a key vector derived from the node embedding. The second term penalizes energy-intensive transitions by subtracting a weighted energy consumption term. We compute
\[
u_{v_t j} = \frac{1}{\sqrt{d_h}} \mathbf{c}_t^\top \mathbf{W}_k \mathbf{z}_j - \lambda \varepsilon_{v_t j},
\]
where $\mathbf{W}_k$ is a learned key projection matrix, $\varepsilon_{v_t j}$ is the normalized energy consumption from the current node to node $j$, and $\lambda > 0$ is a hyperparameter that controls the strength of the energy penalty.

To stabilize training and prevent extreme logit values, we apply hyperbolic tangent clipping to the scores. Specifically, we compute
\[
\tilde{u}_{v_t j} = \kappa \tanh(u_{v_t j} / \kappa),
\]
where $\kappa$ is a clipping parameter. This transformation bounds the range of the scores while preserving their relative ordering. We combine the clipped scores with the feasibility mask to obtain the final action probabilities. If node $j$ violates any constraint, the mask sets $M_t(j) = -\infty$, which ensures that node $j$ receives zero probability after the softmax. The probability of selecting node $j$ as the next action is
\[
\pi_\theta(j \mid s_t) = \frac{\exp(\tilde{u}_{v_t j} + M_t(j))}{\sum_{k \in V} \exp(\tilde{u}_{v_t k} + M_t(k))}.
\]
During training, we sample an action from this distribution to enable exploration. During evaluation, we select the action with the highest probability to obtain the best predicted route.

\subsection{Feasibility Mask}
\label{sec:mask}

This section describes how the constraints from Eq. \eqref{eq:edarp_optimization} are enforced during solution construction through action masking. The feasibility mask ensures that the decoder only considers valid actions at each decision step. At time step $t$, we construct a mask $M_t \in \{0, -\infty\}^N$ that indicates which nodes are allowed. For each candidate next node $j$, we set $M_t(j) = -\infty$ if visiting node $j$ would violate any operational constraint, and $M_t(j) = 0$ otherwise.

We enforce six categories of constraints. First, we prevent revisiting nodes and ensure correct pickup-delivery ordering. We block any node $j$ that has already been served. For delivery nodes, we additionally check that the corresponding pickup has been completed. If node $j$ is a delivery and its paired pickup has not been served, we block node $j$.

Second, we enforce vehicle capacity constraints. Let $l_t$ denote the current passenger load and $q_j$ denote the load change at node $j$. The load change is positive for pickups, negative for deliveries, and zero for the depot and charging stations. We block node $j$ if the resulting load $l_t + q_j$ would fall outside the valid range $[0, Q]$.

Third, we enforce time window constraints. We compute the arrival time at node $j$ as $\tau_t^{\text{arr}} = \tau_t + \delta_{v_t j}$, where $\tau_t$ is the current time and $\delta_{v_t j}$ is the travel time. Service can start at $t_j^{\text{start}} = \max(\tau_t^{\text{arr}}, a_j)$ where $a_j$ is the earliest allowed service time. If we arrive early, we wait until the time window opens. We block node $j$ if $t_j^{\text{start}} > \ell_j$, where $\ell_j$ is the latest allowed service time.

Fourth, we enforce battery constraints. Let $b_t$ denote the current battery level and $\varepsilon_{v_t j}$ denote the energy required to travel to node $j$. We maintain a small safety reserve $\rho$ to prevent complete battery depletion. We block node $j$ if $b_t - \varepsilon_{v_t j} < \rho$. Additionally, we check whether the vehicle can reach the depot or a charging station after visiting node $j$. Let $r_j = \min_{k \in F \cup \{0\}} \varepsilon_{jk}$ denote the minimum energy needed to reach a safe location from node $j$. We block node $j$ if $b_t - \varepsilon_{v_t j} - r_j < \rho$.

Fifth, we enforce operational rules that prevent undesirable behavior and improve solution quality. We block the depot if the vehicle currently carries passengers, as returning with passengers on board would leave requests incomplete. We prevent direct transitions between charging stations to avoid ineffective charging station hopping. We block charging stations immediately after leaving the depot to prevent unnecessary early charging. We also block charging stations when the vehicle carries passengers to avoid inconveniencing customers with charging stops.

Sixth, we manage the vehicle fleet. When a vehicle returns to the depot and unserved requests remain, we check how many vehicles have been used. If fewer than $K$ vehicles have been deployed, we reset the time and battery level to start a new vehicle. If all $K$ vehicles have been used, we block all nodes except the depot to force episode termination.

The computational complexity of constructing the mask is $O(N)$ per decision step, where $N$ is the number of nodes. The time window and battery checks require computing arrival times and energy consumption for each candidate node, which takes constant time per node given precomputed travel times and energy costs. The overall mask construction is dominated by vectorized operations over all candidate nodes.

\subsection{Reward Function}
\label{sec:reward}
The reward function implements the objective from Eq. \eqref{eq:edarp_optimization} within the deep reinforcement learning framework. Let $\mathcal{X}$ denote a complete solution consisting of all vehicle routes. We define three component costs: energy consumption $J_e(\mathcal{X}) = \sum_k \sum_t \varepsilon_{v^k_t v^k_{t+1}}$, waiting time $J_w(\mathcal{X}) = \sum_k \sum_t \max(0, a_{v^k_{t+1}} - \text{arrive}^k_{t+1})$, and lateness at deliveries $J_l(\mathcal{X}) = \sum_k \sum_t \max(0, t^k_{t+1} - b_{v^k_{t+1}}) \mathbb{1}[v^k_{t+1} \in D]$. The total cost combines these components as $J(\mathcal{X}) = w_e J_e(\mathcal{X}) + w_w J_w(\mathcal{X}) + w_l J_l(\mathcal{X}) - w_c n_{\text{served}}(\mathcal{X})$, where $w_e$, $w_w$, $w_l$, and $w_c$ are scalar weights balancing energy, waiting, lateness, and service reward, respectively. The sparse reward $\mathcal{R}(\mathcal{X}) = -J(\mathcal{X})$ is provided only at episode termination, focusing credit assignment on final solution quality rather than intermediate decisions.

\section{Experimental Setup}
\label{sec:experimental_setup}

This section evaluates the proposed approach through two complementary case studies based on real-world ride-sharing data from San Francisco. \textbf{Case Study 1} uses established benchmark instances from the E-DARP literature \citep{BONGIOVANNI2019436} with 16--50 requests in the Civic Center district, enabling direct comparison with exact optimization methods and published results. \textbf{Case Study 2} addresses larger instances with 100--250 requests across a broader San Francisco service area, incorporating realistic energy consumption modeling and more realistic electric vehicle specifications to demonstrate scalability beyond the reach of exact methods.

Both case studies use transportation requests derived from the Uber GPS dataset \citep{uber2011gps}, which contains approximately 1.2 million GPS logs recorded every 4 seconds from active Uber vehicles during one week in San Francisco in 2011. After removing invalid records and extracting pickup and dropoff locations, the processed dataset contains about 25,000 trips with daily request volumes ranging from 2,000 to 6,000.

\subsection{Case Study 1: Benchmark Instances}
\label{sec:case_study_1}

The first case study uses benchmark instances introduced by \citep{BONGIOVANNI2019436} to enable direct comparison with published results. These instances have been widely adopted in the E-DARP literature, allowing assessment of how close learned policies come to optimal or best-known solutions.

The benchmark instances use transportation requests from the Uber GPS dataset \citep{uber2011gps} for San Francisco's Civic Center area. The transportation network is extracted from OpenStreetMap, and charging station locations are obtained from the Alternative Fuels Data Center of the U.S. Department of Energy. Instance sizes range from 16 to 50 requests served by 2 to 5 vehicles. The benchmark defines five depot locations that coincide with five charging station locations, allowing vehicles to start, end, and recharge at the same facilities. The original benchmark considers three minimum state-of-charge thresholds (0.1, 0.4, and 0.7) to study battery management under different safety margins. We evaluate only the 0.1 threshold, which represents the most realistic operational scenario with minimal safety buffer.

The benchmark objective differs from the formulation in Section \ref{sec:problem_statement}. Rather than minimizing energy, waiting time, and lateness, the benchmark minimizes travel time and excess travel time with weights $w_1 = 0.75$ for travel time and $w_2 = 0.25$ for excess travel time. Additionally, the benchmark formulation does not include the completion bonus term, implicitly assuming all requests must be served. We adopt these objective weights and constraint specifications to ensure direct comparability with published results.

A key difference between the benchmark formulation and our learned approach concerns the treatment of charging decisions. In the original benchmark, charging time at each station visit is a continuous decision variable, allowing the optimization to determine the exact amount of energy to charge based on future route requirements \citep{BONGIOVANNI2019436}. Our reinforcement learning framework, however, requires a fixed charging time or charging threshold and cannot directly optimize continuous charging durations within the policy network. To address this, we fix the charging time to a predetermined duration at each charging station visit. Setting a relatively small fixed charging time provides finer-grained control to the learned policy: the agent can effectively modulate total charging by deciding how many times to visit charging stations, including consecutive visits to the same station, rather than how long to charge per visit. This discretization trades some solution flexibility for compatibility with our action-based learning framework while still enabling effective battery management strategies. However, this approach may yield suboptimal solutions in scenarios where charging is required, as the fixed duration cannot adapt to the precise energy needs of each route, potentially resulting in either insufficient charge or unnecessary time spent at stations.

The two case studies use different objective functions to address distinct evaluation goals. Case Study 1 adopts the benchmark objective to enable direct comparison with published exact methods. This isolates the question of whether learned policies can match optimization approaches on established instances. Case Study 2 uses the full E-DARP objective from Section \ref{sec:reward}, which better reflects operational priorities for electric fleets such as energy efficiency and service completion under battery constraints. Although this prevents direct numerical comparison between case studies, both objectives share the core structure of minimizing travel costs subject to time window and capacity constraints. The strong performance across both formulations suggests the GREAT architecture captures the underlying routing structure rather than being specific to one objective.

Training instances for Case Study 1 are generated following these same specifications, sampling requests from the processed Uber GPS dataset. The neural network is trained using REINFORCE \cite{williams1992simple} with POMO  \cite{kwon2020pomo} to increase variance and improve solution quality. To evaluate generalization across problem scales, we employ mixed-size training with instances containing $n \in \{16, 24, 32, 40, 50\}$ request pairs, sampled uniformly during training. The training dataset consists of 5000 graphs, the validation dataset of 1000 graphs, and the test dataset of the benchmark cases.

\subsection{Case Study 2: Large-Scale Realistic Instances}
\label{sec:case_study_2}
The second case study focuses on larger problem instances ranging from 100 to 250 request pairs, representing more realistic operational scenarios for urban ride-sharing services. While not yet approaching the thousands of daily trips characteristic of full-scale deployments, these instances incorporate realistic energy consumption modeling, expanded service areas, and modern electric vehicle specifications that better reflect practical fleet operations. This scale remains tractable for evaluation and analysis while demonstrating the approach's viability for problems where exact optimization methods become computationally prohibitive.

Transportation requests are drawn from the Uber GPS dataset \citep{uber2011gps} covering a broader region of San Francisco beyond the Civic Center area. The transportation network is extracted from OpenStreetMap \citep{OpenStreetMap}, providing detailed road topology and connectivity. This network is augmented with elevation data from OpenTopography \citep{opentopography} to account for San Francisco's hilly terrain. Energy consumption is computed using a physics-based model that considers vehicle speed, acceleration, road grade, and powertrain efficiency, following the approach of \citep{basso2019energy}. The model accounts for both traction energy requirements and regenerative braking on downhill segments, providing substantially more realistic energy dynamics than the simplified linear models in Case Study 1.

Fleet configurations represent modern electric vehicle specifications with battery capacities of 20 and 40 kWh, fleet sizes of 4 and 8 vehicles, and ride-sharing capacities of 1, 2, and 3 passengers. Charging dynamics follow the nonlinear charging curve defined in Eq. \eqref{eq:charging_power}, based on empirical data from \citep{wassiliadis2021review}. Ten charging stations are distributed across the city, with locations determined by K-means clustering on the transportation network nodes. Time windows are set to 15 minutes for both pickups and deliveries, with delivery windows offset from pickup windows by the direct travel time between origin and destination. This provides moderate flexibility while maintaining realistic service quality requirements.

The objective function uses the full formulation from Section \ref{sec:problem_statement}, incorporating energy consumption ($w_e$), waiting time ($w_w$), lateness ($w_l$), and completion bonus ($w_c$) with weights balanced to reflect operational priorities. Unlike the benchmark instances in Case Study 1 where all requests must be served, this formulation explicitly accounts for resource-constrained scenarios where battery capacity or fleet size may be insufficient to complete all demand.

For training on larger instances in Case Study 2, curriculum learning is employed to address computational challenges. Training directly on large instances is slow due to problem complexity and GPU memory constraints that force smaller batch sizes. Training begins with a model trained on $n=100$ request pairs and progressively fine-tunes it on larger instances, increasing by 20\% at each step: [100, 120, 144, 173, 207, 250]. For each instance size, a dataset of 2,000 instances is created and the model is fine-tuned for five epochs with POMO rollout of 100 trajectories. This few-shot approach leverages knowledge from the base model to adapt efficiently to larger problem scales without the computational expense of training from scratch.

For both case studies, the Adam optimizer \citep{kingma2015adam} is used  and POMO augmentation is applied by generating multiple parallel rollouts from different initial conditions for each training instance. All training and inference experiments were conducted on the Alvis cluster, a national computing resource provided by the Swedish National Academic Infrastructure for Supercomputing (NAISS) and dedicated to artificial intelligence and machine learning research. Experiments were performed using NVIDIA A40 GPUs. This hardware configuration was used consistently across all reported experiments to ensure fair comparison of computational performance.

\subsection{Baseline Methods}
\label{sec:baselines}

The proposed approach is compared against several baseline methods representing different algorithmic paradigms. Specifically, we compare against: (1) a greedy nearest-neighbor heuristic \citep{rosenkrantz1977analysis}, which establishes a lower performance bound and quantifies gains from learned policies over myopic decision-making; (2) Adaptive Large Neighborhood Search (ALNS), the dominant metaheuristic paradigm for vehicle routing problems \citep{ropke2006adaptive}, providing a strong optimization baseline with practical computational requirements; and (3) exact methods including branch-and-price and branch-and-cut, which establish upper bounds on achievable solution quality. Together, these baselines span the spectrum from fast construction heuristics to computationally intensive exact solvers, enabling comprehensive evaluation of both solution quality and computational efficiency.

\paragraph{Greedy heuristic.}
We include a greedy heuristic to establish a lower performance bound and quantify the benefit of learned policies over myopic, rule-based decision-making.
A nearest-neighbor greedy heuristic is implemented that constructs routes by iteratively selecting the closest feasible node from the current location. At each decision point, the heuristic evaluates all feasible nodes (those satisfying capacity, battery, time window, and precedence constraints) and selects the one with minimum energy consumption. When the current vehicle cannot feasibly serve any remaining requests, it returns to the depot and a new vehicle is initialized if the fleet limit allows.

\paragraph{ALNS.} Adaptive Large Neighborhood Search represents the state-of-the-art 
metaheuristic paradigm for vehicle routing problems \cite{ropke2006adaptive}, providing 
a strong optimization baseline that balances solution quality with practical computational 
requirements. ALNS has been successfully applied to dial-a-ride variants, including static 
DARP with service quality objectives \cite{pfeiffer2022alns} and dynamic E-DARP with 
predictive routing \cite{bongiovanni2020predictive}. We adapt the ALNS framework for our 
static E-DARP setting, incorporating destroy and repair operators suited to the 
pickup-delivery structure and battery constraints.

We benchmark our DRL policy against an ALNS heuristic with a maximum runtime of 2 hours 
per instance. ALNS starts from a feasible greedy rollout and iteratively improves the 
solution through destroy-repair cycles. In each iteration, a subset of pickup-delivery 
requests is removed from the current route and reinserted to reconstruct a complete 
solution. Feasibility and objective evaluation are performed by replaying candidate 
routes in the E-DARP environment, which enforces time windows, precedence, capacity, 
and battery constraints. The objective matches the E-DARP reward: a weighted sum of 
energy, waiting time, and late delivery penalties, plus completion reward.

We employ three destroy operators: random removal, which uniformly samples requests 
for deletion; Shaw removal \cite{shaw1998using}, which removes clusters of related 
requests based on pickup and delivery travel times and time-window similarity; and 
worst removal, which targets requests whose removal yields the largest marginal 
improvement. For reconstruction, we use random insertion as well as regret-$k$ insertion 
with $k \in \{2, 3\}$, where requests with the largest gap between their best and 
$k$-th best insertion position are prioritized \cite{pfeiffer2022alns}.

Operators are selected via roulette-wheel sampling with adaptive weights updated every 
segment, following the weight adaptation scheme of \cite{ropke2006adaptive}. We use 
rewards $\sigma_1 = 10$, $\sigma_2 = 5$, and $\sigma_3 = 1$ for discovering a new 
global best, improving the current solution, and accepting a non-improving solution, 
respectively. Weights are updated using exponential smoothing with decay rate 
$\rho = 0.2$, and we enforce a minimum weight to preserve exploration.

For solution acceptance, we use record-to-record travel (RTR) \cite{dueck1993new}, which accepts any 
solution within a tolerance of the best found so far. The initial tolerance is set 
to 5\% of the initial solution cost and decays by a factor of 0.99 per iteration.

\paragraph{Exact methods (Case Study 1 only).}
Exact methods establish an upper bound on achievable solution quality, enabling precise quantification of the optimality gap of learned policies on standardized benchmarks.
For Case Study 1, additional comparison is made against the results reported in \citep{su2024branch}, which include exact solutions obtained through column generation (CG), branch-and-price (B\&P), and branch-and-cut (B\&C) for smaller instances, and best-known solutions from advanced optimization techniques for larger instances.

\section{Results and Discussion}
\label{sec:results}

\subsection{Case Study 1: Benchmark Performance}
\label{sec:results_case1}

This case study evaluates solution quality and computational efficiency on standard benchmark instances, comparing GREAT against four exact optimization methods: column generation (CG),column generation with cutting planes (CG-CP), branch-and-price (B\&P), and branch-and-cut (B\&C) from \citep{su2024branch}. Table \ref{tab:results_case1} presents solution quality gaps relative to best-known solutions alongside computation times.

\begin{table}
    \centering
    \caption{Solution quality and computation time comparison on benchmark instances. Methods compared: Column Generation (CG), Column Generation with cutting planes (CG-CP), Branch \& Price (B\&P), Branch \& Cut (B\&C), and GREAT. Gap computed as percentage deviation from best-known solution: $\text{Gap}\% = 100 \times (z - z^*)/z^*$. Exact methods terminated at 7,200-second limit for u5-50.}
    \label{tab:results_case1}
    \begin{tabular}{lcccccccccc}
    \toprule
        & \multicolumn{2}{c}{\textbf{CG}} & \multicolumn{2}{c}{\textbf{CG-CP}} & \multicolumn{2}{c}{\textbf{B\&P}} & \multicolumn{2}{c}{\textbf{B\&C}} & \multicolumn{2}{c}{\textbf{GREAT}} \\
        \cmidrule(lr){2-3} \cmidrule(lr){4-5} \cmidrule(lr){6-7} \cmidrule(lr){8-9} \cmidrule(lr){10-11}
        \textbf{Instance} & Gap\% & Time (s) & Gap\% & Time (s) & Gap\% & Time (s) & Gap\% & Time (s) & Gap\% & Time (s) \\ 
    \midrule
        u2-16 (16 requests) & 0.92 & 43.6 & 0.00 & 72.1 & 0.00 & 22.2 & 0.00 & 21.0 & 0.00 & $<$1.0 \\ 
        u3-30 (30 requests) & 0.00 & 570.8 & 0.00 & 570.8 & 0.00 & 570.8 & 0.00 & 438.0 & 0.00 & $<$1.0 \\ 
        u5-50 (50 requests) & 0.05 & 7200.0$^{\dagger}$ & 0.05 & 7200 & 0.05 & 7200.0$^{\dagger}$ & 7.09 & 7200.0$^{\dagger}$ & 0.40 & $<$1.0 \\ 
    \bottomrule
    \multicolumn{11}{l}{\footnotesize $^{\dagger}$Time limit reached without proven optimality} \\
    \end{tabular}
\end{table}

GREAT achieves optimal or near-optimal solutions across all instances while providing substantial computational advantages. For u2-16, GREAT matches the proven optimal solutions found by CG-CP, B\&P, and B\&C, whereas CG exhibits a 0.92\% gap. All methods find optimal solutions for u3-30, demonstrating that the learned policy captures essential problem structure at this scale.

For u5-50, GREAT achieves a 0.40\% gap compared to 0.05\% for CG, CG-CP, and B\&P. While these exact methods yield marginally better solutions, B\&C struggles with a 7.09\% gap. GREAT thus outperforms B\&C by 6.69 percentage points, indicating effective learning of routing and charging strategies. Importantly, the exact methods reached the two-hour time limit without proving optimality, whereas GREAT produces its solution in sub-second time.

The computational advantage of GREAT increases with problem size. For u2-16, GREAT achieves a 21$\times$ to 44$\times$ speedup over exact methods. For u3-30, this advantage grows to over 400$\times$, with GREAT responding in under one second compared to approximately ten minutes for exact methods. For u5-50, all exact methods reach the 7,200-second limit while GREAT maintains sub-second inference, representing over 7,200$\times$ speedup at 0.40\% optimality gap.

These results demonstrate that the learned policy scales more favorably than exact methods, which exhibit exponential growth in computation time. The consistent sub-second response time across all instance sizes confirms computational stability essential for real-time deployment. The trade-off of a small optimality gap for several orders of magnitude speedup represents a practical compromise suitable for dynamic operational environments.

\subsection{Case Study 2: Scalability Analysis}
\label{sec:results_case2}


To assess the scalability of our approach to larger problem instances, we evaluate the curriculum-trained reinforcement learning policy on a challenging test scenario of 250 requests and 10 electric vehicles. The policy was progressively trained from 100 to 250 request pairs. We compare against two established baseline methods: a greedy nearest-neighbor heuristic that myopically assigns vehicles to the closest feasible requests, and an Adaptive Large Neighborhood Search (ALNS) metaheuristic with a computational budget of 2 hours per instance. ALNS is a state-of-the-art optimization method widely used for vehicle routing problems, combining multiple destroy and repair operators within a simulated annealing framework. Table \ref{tab:results} presents the comprehensive performance comparison across solution quality, fleet utilization, service quality, and computational efficiency metrics.

\begin{table}
\centering
\caption{Performance comparison on E-DARP test instances (250 requests, 10 vehicles, averaged over 20 graphs). Best results are shown in bold.}
\label{tab:results}
\begin{tabular}{llccc}
\toprule
\textbf{Category} & \textbf{Metric} & \textbf{Greedy} & \textbf{ALNS} & \textbf{GREAT (Ours)} \\
\midrule
\multirow{3}{*}{Solution Quality} 
& Test Profit & 209.45 & 847.07 & \textbf{927.25} \\
& Completion Rate (\%) & 24.0 & 94.6 & \textbf{100.0} \\
& Requests Served & 60/250 & 236.5/250 & \textbf{250/250} \\
\midrule
\multirow{4}{*}{Fleet Utilization}
& Vehicles Used & 10.0 & 9.4 & \textbf{8.25} \\
& Load Factor & 1 & 1 & \textbf{1.94} \\
& Charge Visits & 11.8 & 1.25 & \textbf{0.45} \\
& Energy per Vehicle (kWh) &\textbf{3.2} & 11.6 & 9.6 \\
\midrule
\multirow{2}{*}{Service Quality}
& Wait per Request (s) & 168.6 & \textbf{8.3} & 33.2 \\
& Lateness per Request (s) & 35.4 & \textbf{0.0} & 30.3 \\
\midrule
Computational
& Computation Time (s) & 3.0 & 7200.0 & \textbf{9.3} \\
\bottomrule
\end{tabular}
\end{table}

Our RL approach achieves the highest test profit of 927.25, outperforming ALNS by 9.5\% (847.07) and greedy by 343\% (209.45). Most notably, our method is the only one to achieve 100\% completion rate, successfully serving all 250 requests, whereas ALNS achieves 94.6\% completion (236.5 requests) and the greedy heuristic serves only 24\% of requests (60 out of 250). This demonstrates that our learned policy effectively balances the complex trade-offs inherent in the E-DARP, including time window constraints, vehicle capacity, and battery management.

The RL approach demonstrates superior fleet utilization, requiring only 8.25 vehicles on average compared to 9.4 for ALNS and the full fleet of 10 for the greedy baseline. This 12\% reduction in vehicles used compared to ALNS translates directly to operational cost savings. Furthermore, our method achieves a load factor of 1.94, indicating effective ridesharing with nearly two passengers transported simultaneously on average, whereas both ALNS and greedy achieve a load factor of 1, indicating single-passenger trips that underutilize vehicle capacity. The RL policy also learns efficient battery management, requiring only 0.45 charging station visits per route compared to 1.25 for ALNS and 11.8 for greedy. Despite serving all requests, RL consumes only 9.6 kWh per vehicle compared to 11.6 kWh for ALNS, demonstrating more energy-efficient routing. The greedy heuristic uses just 3.2 kWh per vehicle, but this low energy consumption reflects its poor completion rate rather than efficiency.

ALNS achieves the lowest wait and lateness times per request. However, this comes at the cost of serving fewer customers. ALNS effectively selects easier requests that can be served with minimal delay, leaving 13.5 requests (5.4\%) unserved. Our RL approach prioritizes serving all customers and maximizing ride-sharing, accepting slightly higher average wait times (33.2s vs 8.3s) and lateness (30.3s vs 0.0s) as a trade-off for 100\% completion and nearly 2$\times$ higher vehicle utilization. Crucially, our method achieves the highest overall profit, demonstrating that the gains from complete service coverage and efficient fleet utilization outweigh the modest increase in lateness under the E-DARP objective.

Perhaps the most striking advantage of our approach is computational efficiency. The RL policy generates solutions in approximately 9.3 seconds per instance, representing a 775$\times$ speedup compared to ALNS which requires 7,200 seconds (2 hours) per instance. This dramatic difference stems from the fundamental nature of the approaches: ALNS performs iterative local search requiring thousands of solution evaluations, while our trained neural network constructs solutions through sequential forward passes. While the greedy heuristic is faster at 3.0 seconds per instance, it achieves only 24\% completion rate, demonstrating that the modest additional computation time of our approach yields substantial gains in solution quality. This computational efficiency has significant practical implications for dynamic ride-sharing scenarios where customer requests arrive throughout the day, as the ability to recompute solutions in seconds rather than hours enables responsive fleet management and practical operational deployment.

The strong scalability of the learned policy can be attributed to several architectural and methodological factors. First, the GREAT encoder's edge-based attention mechanism directly models the routing decision space, learning which transitions are promising based on relational features rather than absolute positions. These learned patterns, such as prioritizing requests with tight time windows or balancing vehicle workloads, generalize across problem sizes because the underlying decision logic remains consistent regardless of instance scale. Second, curriculum learning enables the policy to progressively build competence on increasingly complex instances, transferring knowledge from smaller problems rather than learning large-scale coordination from scratch. Finally, inference requires only a single forward pass through the network, yielding polynomial-time complexity compared to the exponential growth in search space faced by exact methods and the iterative evaluation required by ALNS.

\subsection{Hyperparameter Sensitivity Analysis}
A comprehensive hyperparameter sensitivity analysis was conducted to identify the optimal neural architecture. In total, 10 configurations were evaluated by varying three key hyperparameters: hidden dimension $d \in \{64, 128, 192\}$, number of attention heads $h \in \{4, 8, 12\}$, and number of transformer layers $L \in \{4, 5, 6\}$. Experiments used an 8-vehicle fleet with 20 kWh batteries and 3-passenger capacity.

\begin{figure}
    \centering
    \includegraphics[width=0.7\textwidth]{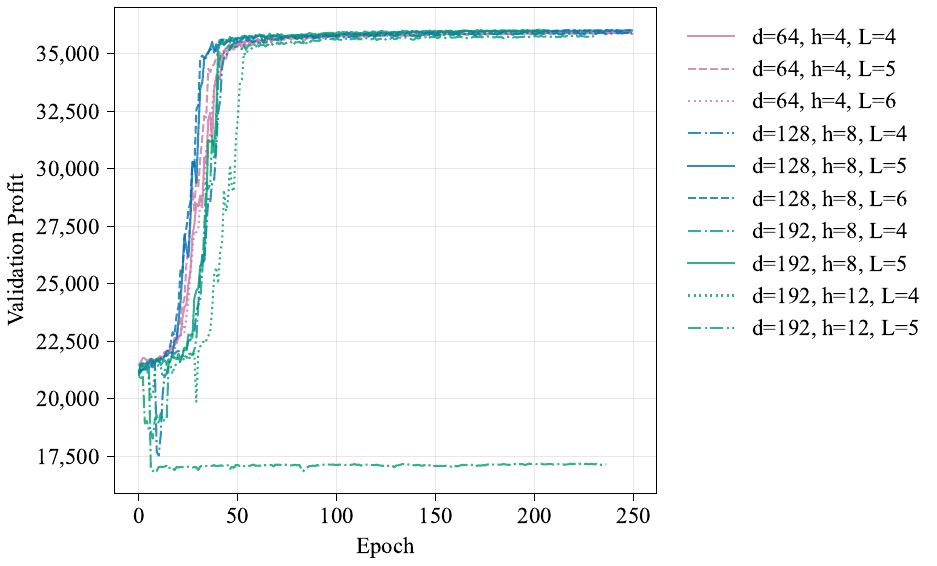}
    \caption{Validation profit over training epochs for all hyperparameter configurations. Most configurations converge to similar performance levels ($\sim$36000) by epoch 50, with $d=128$ (blue) and $d=64$ (purple) configurations showing comparable convergence rates. The $d=192$, $h=8$, $L=5$ configuration (green dashed) diverged early and failed to recover.}
    \label{fig:val_profit}
\end{figure}

The best performing configuration achieved a maximum validation profit of 36048 at epoch 203 with $d=192$, $h=8$, $L=6$, representing a \textbf{67.3\%} improvement over the worst configuration, Table \ref{tab:hyperparameter_results}. However, $d=128$ configurations offer an attractive middle ground: the $d=128$, $h=8$, $L=5$ variant achieves competitive performance (35952 max profit, only 0.3\% below the best) while converging quickly (95\% of peak by epoch 36) and providing faster inference speed compared to $d=192$ models. Fig. \ref{fig:val_profit} shows the validation profit curves for all configurations, illustrating the convergence behavior across different hyperparameter settings. To quantify training stability, we report $\sigma_{50}$, the standard deviation of validation profit over the last 50 epochs, lower values indicate more consistent performance during the final training phase. Most configurations showed excellent stability with $\sigma_{50}$ below 25. Notably, one configuration ($d=192$, $h=8$, $L=5$) failed to converge, plateauing at a significantly lower profit level.

\begin{table}
\centering
\caption{Hyperparameter sensitivity analysis results. Configurations are ordered by decreasing maximum validation profit. Best configuration shown in \textbf{bold}. $d$: hidden dimension, $h$: attention heads, $L$: transformer layers. Ep$_{95\%}$: epochs to reach 95\% of max profit (convergence speed). $\sigma_{50}$: standard deviation over last 50 epochs (stability). Improv.: gain relative to worst configuration.}

\label{tab:hyperparameter_results}
\begin{tabular}{ccc|rrrrrr}
\toprule
$d$ & $h$ & $L$ & Max Profit & Epoch & Final Profit & Ep$_{95\%}$ & $\sigma_{50}$ & Improv. (\%) \\
\midrule
\textbf{192} & \textbf{8} & \textbf{6} & \textbf{36048} & \textbf{203} & \textbf{36019} & \textbf{41} & \textbf{23.1} & \textbf{67.3} \\
192 & 8 & 4 & 36031 & 218 & 36019 & 32 & 21.1 & 67.2 \\
64 & 4 & 5 & 36026 & 221 & 36009 & 34 & 18.9 & 67.2 \\
64 & 4 & 4 & 35969 & 234 & 35954 & 41 & 22.7 & 66.9 \\
128 & 8 & 6 & 35956 & 219 & 35928 & 54 & 22.2 & 66.9 \\
128 & 8 & 5 & 35952 & 249 & 35934 & 36 & 21.5 & 66.8 \\
128 & 8 & 4 & 35921 & 244 & 35856 & 43 & 19.1 & 66.7 \\
64 & 4 & 6 & 35920 & 235 & 35914 & 43 & 22.6 & 66.7 \\
192 & 12 & 4 & 35793 & 229 & 35768 & 41 & 22.8 & 66.1 \\
192 & 8 & 5 & 21548 & 6 & 17143 & 1 & 15.8 & 0.0 \\
\bottomrule
\end{tabular}
\end{table}

\subsection{Sensitivity Analysis: Reward Function Weights}

\begin{figure*}
    \centering
    \begin{subfigure}[b]{0.32\textwidth}
        \centering
        \includegraphics[width=\textwidth]{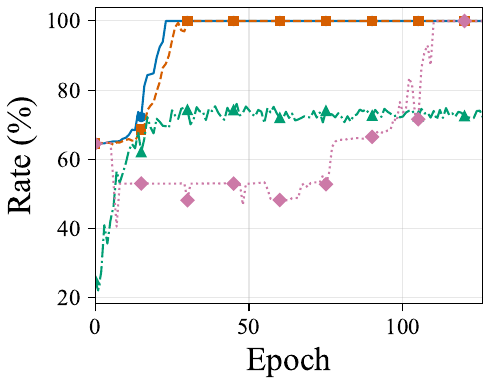}
        \caption{Completion Rate}
    \end{subfigure}
    \hfill
    \begin{subfigure}[b]{0.32\textwidth}
        \centering
        \includegraphics[width=\textwidth]{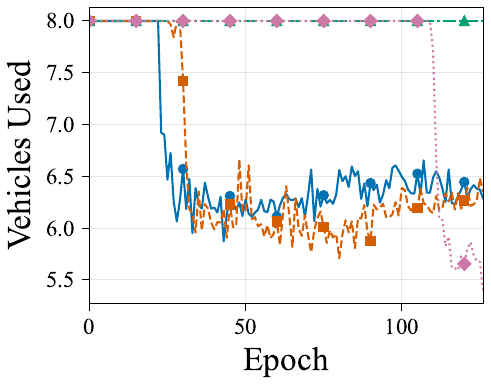}
        \caption{Active Vehicles}
    \end{subfigure}
    \hfill
    \begin{subfigure}[b]{0.32\textwidth}
        \centering
        \includegraphics[width=\textwidth]{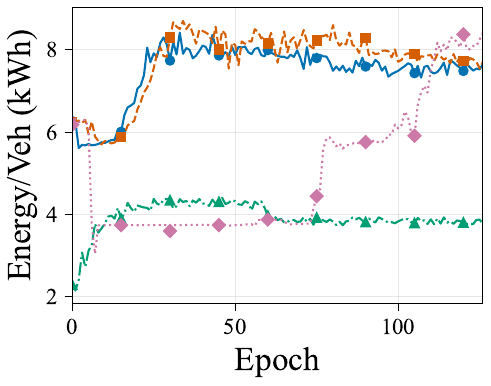}
        \caption{Energy Consumption}
    \end{subfigure}
    
    \vspace{0.5em}
    
    \begin{subfigure}[b]{0.32\textwidth}
        \centering
        \includegraphics[width=\textwidth]{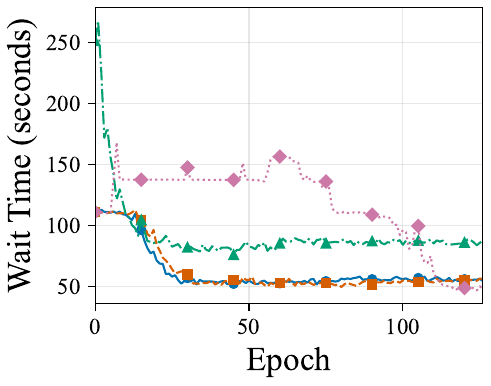}
        \caption{Service Wait Time}
    \end{subfigure}
    \hfill
    \begin{subfigure}[b]{0.32\textwidth}
        \centering
        \includegraphics[width=\textwidth]{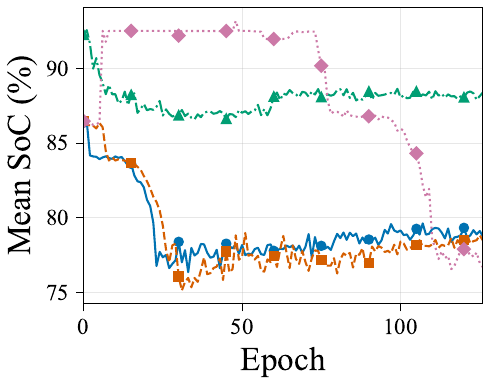}
        \caption{State of Charge}
    \end{subfigure}
    \hfill
    \begin{subfigure}[b]{0.32\textwidth}
        \centering
        \includegraphics[width=\textwidth]{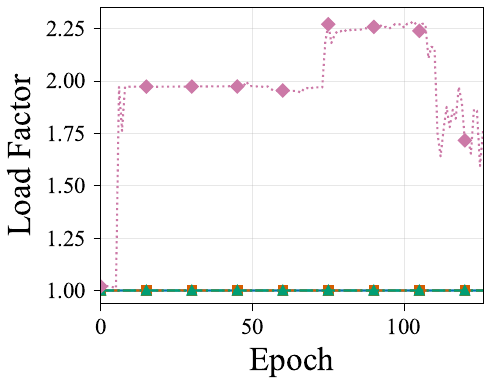}
        \caption{Vehicle Load Factor}
    \end{subfigure}
    
    \vspace{0.5em}
    
    \includegraphics[width=0.9\textwidth]{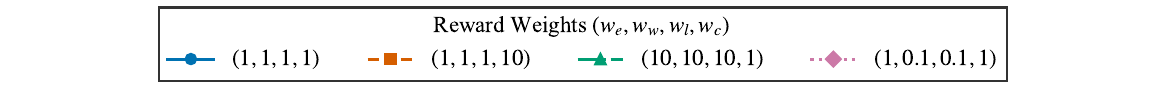}
    
    \caption{Training dynamics across reward weight configurations $(w_e, w_w, w_l, w_c)$ for 100 requests with 20 kWh batteries and 8 vehicles. Metrics shown: (a) completion rate, (b)nvehicles deployed, (c)energy consumption per vehicle, (d) customer wait time, (e) mean state of charge, and (f) vehicle load factor. High operational cost weights $(10,10,10,1)$ lead to conservative behavior and incomplete service, while reduced service penalties $(1,0.1,0.1,1)$ enable ride-sharing (load factor $>1.8$) at the cost of moderate lateness.}
    \label{fig:reward_sensitivity}
\end{figure*}

This subsection presents a sensitivity analysis examining how different reward function weight configurations affect system performance. The scenario is for 100 requests, 20 kWh battery and 8 vehicles. The reward function weights $(w_e, w_w, w_l, w_c)$ control the relative importance of energy consumption, wait time, lateness, and completion penalty terms respectively. Fig. \ref{fig:reward_sensitivity} presents the training dynamics across all configurations, and Table \ref{tab:reward_sensitivity} summarizes key operational and service quality metrics.

\begin{table*}
\centering
\caption{Reward weight sensitivity analysis: operational and service quality metrics. Weights are specified as $(w_e, w_w, w_l, w_c)$ representing energy, wait time, lateness, and completion penalties respectively.}
\label{tab:reward_sensitivity}
\footnotesize
\begin{tabular}{lcccccccc}
\toprule
\textbf{Reward Weights} & \textbf{Completion} & \textbf{Vehicles} & \textbf{Load} & \textbf{Energy/Veh} & \textbf{Total Energy} & \textbf{Mean SoC} & \textbf{Wait Time} & \textbf{Lateness} \\
$(w_e, w_w, w_l, w_c)$ & \textbf{Rate (\%)} & \textbf{Used} & \textbf{Factor} & \textbf{(kWh)} & \textbf{(kWh)} & \textbf{(\%)} & \textbf{(s)} & \textbf{(s)} \\
\midrule
(1, 1, 1, 1) & 100.0 & 6.11 & 1.00 & 7.10 & 43.4 & 75.1 & 57.6 & 0.0 \\
(1, 1, 1, 10) & 100.0 & 6.15 & 1.00 & 7.80 & 48.0 & 76.8 & 57.0 & 0.0 \\
(10, 10, 10, 1) & 74.0 & 8.00 & 1.00 & 3.80 & 30.4 & 86.9 & 64.2 & 0.0 \\
(1, 0.1, 0.1, 1) & 100.0 & 5.61 & 1.82 & 8.30 & 46.6 & 76.8 & 48.9 & 39.7 \\
\bottomrule
\end{tabular}
\end{table*}

The sensitivity analysis reveals important trade-offs between completion incentives and operational cost penalties. As shown in Fig. \ref{fig:reward_sensitivity}(a), the baseline configuration $(1, 1, 1, 1)$ with equal weights achieves 100\% completion rate, utilizing 6.11 vehicles with an energy consumption of 7.10 kWh per vehicle and average wait time of 58 seconds per request.

Increasing the completion weight tenfold $(1, 1, 1, 10)$ also achieves 100\% completion. This configuration uses slightly more vehicles (6.15) with higher energy consumption (7.80 kWh) and comparable wait times (57 seconds). The stronger completion incentive encourages the policy to prioritize service fulfillment, resulting in slightly higher resource utilization as evident in Fig. \ref{fig:reward_sensitivity}(b).

However, scaling operational cost weights $(10, 10, 10, 1)$ while maintaining unit completion weight produces suboptimal results. This configuration achieves only 74.0\% completion rate despite utilizing all 8 vehicles. As illustrated in Fig. \ref{fig:reward_sensitivity}(c,e), the model exhibits significantly lower energy consumption (3.80 kWh per vehicle) and higher mean state-of-charge (86.9\%), with longer wait times (64 seconds). The high operational penalties cause overly conservative behavior, with the policy prioritizing energy conservation and minimizing wait/lateness penalties over completing requests.

The configuration $(1, 0.1, 0.1, 1)$ demonstrates an alternative trade-off by reducing the weights on wait time and lateness while maintaining equal emphasis on energy and completion. This configuration achieves 100\% completion rate with the fewest vehicles (5.61) and a load factor of 1.82, compared to 1.0 for all other configurations. This difference is significant: a load factor of 1.0 indicates single-occupancy trips with no ride-sharing, while a load factor of 1.82 demonstrates that the policy has learned to pool multiple passengers per vehicle trip. The ride-sharing behavior results in lower average wait times (49 seconds vs.\ 57--64 seconds) due to more efficient vehicle utilization, but also leads to non-zero lateness (40 seconds mean) as vehicles make detours to serve multiple passengers. While the energy consumption per vehicle (8.30 kWh) is higher than the equal-weight configurations due to the increased workload from ride-sharing, the total fleet energy consumption (46.6 kWh) remains comparable to the baseline (43.4 kWh) and lower than the high-completion configuration (48.0 kWh).

These results indicate that adequate completion weight is essential for achieving full service coverage. The baseline $(1, 1, 1, 1)$ offers balanced operation with zero lateness but no ride-sharing, while the $(1, 0.1, 0.1, 1)$ configuration enables ride-sharing behavior through reduced wait and lateness penalties. We adopt the $(1, 0.1, 0.1, 1)$ configuration for subsequent experiments as it achieves full completion with efficient fleet utilization through ride-sharing, and the modest lateness trade-off is acceptable for the ride-sharing scenarios considered.

\subsection{Sensitivity Analysis: Battery Size, Fleet Size and Ride-Sharing Capacity}
\label{sec:sensitivity}

\begin{figure*}
\centering
\begin{subfigure}[b]{0.48\textwidth}
    \centering
    \includegraphics[width=\textwidth]{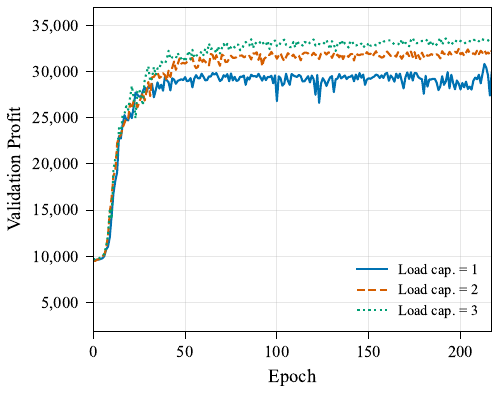}
    \caption{20 kWh battery, 4 vehicles}
    \label{fig:convergence_bat20_fleet4}
\end{subfigure}
\hfill
\begin{subfigure}[b]{0.48\textwidth}
    \centering
    \includegraphics[width=\textwidth]{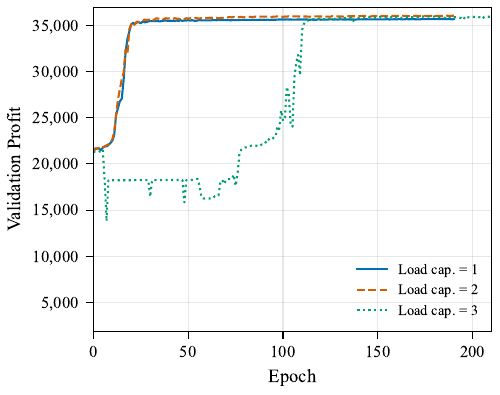}
    \caption{20 kWh battery, 8 vehicles}
    \label{fig:convergence_bat20_fleet8}
\end{subfigure}
\vspace{0.5em}
\begin{subfigure}[b]{0.48\textwidth}
    \centering
    \includegraphics[width=\textwidth]{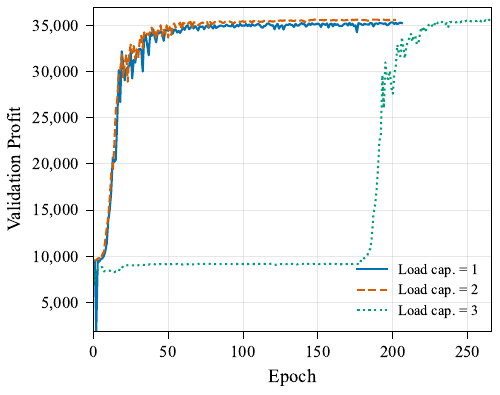}
    \caption{40 kWh battery, 4 vehicles}
    \label{fig:convergence_bat40_fleet4}
\end{subfigure}
\hfill
\begin{subfigure}[b]{0.48\textwidth}
    \centering
    \includegraphics[width=\textwidth]{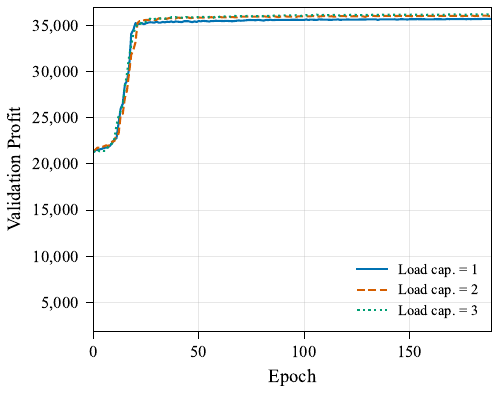}
    \caption{40 kWh battery, 8 vehicles}
    \label{fig:convergence_bat40_fleet8}
\end{subfigure}
\caption{Validation profit convergence trajectories for all twelve configurations. Each subplot shows three ride-sharing capacity settings (1, 2, and 3 passengers per vehicle) for a given battery and fleet size combination.}
\label{fig:convergence_all}
\end{figure*}

This subsection examines how battery capacity, fleet size, and ride-sharing capacity affect system performance. We conduct the analysis on a 100-request scenario with 10 charging stations and 1 depot. The analysis considers battery capacities of 20 and 40 kWh, fleet sizes of 4 and 8 vehicles, and ride-sharing capacities of 1, 2, and 3 passengers per vehicle, yielding twelve configurations in total. Fig. \ref{fig:convergence_all} presents validation profit trajectories for all configurations, while Table \ref{tab:sensitivity_analysis} summarizes key operational metrics at best-performing epochs and Table \ref{tab:sensitivity_request_stats} presents per-request service quality metrics.

\begin{table*}
\centering
\caption{Sensitivity analysis results showing operational metrics at best-performing epochs. Bold values indicate best performance within each category.}
\label{tab:sensitivity_analysis}
\footnotesize
\begin{tabular}{cccccccccc}
\toprule
\textbf{Battery} & \textbf{Fleet} & \textbf{Load} & \textbf{Val.} & \textbf{Vehicles} & \textbf{Energy/} & \textbf{Total} & \textbf{Completion} & \textbf{Charge} & \textbf{Load} \\
\textbf{(kWh)} & \textbf{Size} & \textbf{Cap.} & \textbf{Profit} & \textbf{Used} & \textbf{Vehicle (kWh)} & \textbf{Energy (kWh)} & \textbf{Rate (\%)} & \textbf{Visits} & \textbf{Factor} \\
\midrule
20 & 4 & 1 & 30,824 & 4.00 & 9.05 & 36.20 & 85.5 & 0.00 & 1.00 \\
20 & 4 & 2 & 32,473 & 4.00 & 9.30 & 37.20 & 89.9 & 0.00 & 1.34 \\
20 & 4 & 3 & 33,652 & 4.00 & 9.55 & 38.20 & 93.2 & 0.00 & 1.90 \\
20 & 8 & 1 & 35,699 & 6.04 & 7.85 & 47.43 & \textbf{100.0} & 1.82 & 1.00 \\
20 & 8 & 2 & 36,054 & 5.66 & \textbf{7.60} & 43.02 & \textbf{100.0} & 1.52 & 1.36 \\
20 & 8 & 3 & 35,986 & 5.07 & 8.60 & 43.60 & \textbf{100.0} & 0.00 & \textbf{1.80} \\
40 & 4 & 1 & 35,340 & 3.92 & 12.76 & 50.03 & 99.8 & 0.00 & 1.00 \\
40 & 4 & 2 & 35,657 & 3.65 & 12.93 & 47.18 & \textbf{100.0} & 0.00 & 1.30 \\
40 & 4 & 3 & 35,670 & \textbf{3.58} & 13.07 & 46.80 & \textbf{100.0} & 3.01 & 1.78 \\
40 & 8 & 1 & 35,715 & 5.85 & 8.07 & 47.20 & \textbf{100.0} & 1.65 & 1.00 \\
40 & 8 & 2 & 36,066 & 5.36 & \textbf{8.06} & 43.22 & \textbf{100.0} & \textbf{0.59} & 1.33 \\
40 & 8 & 3 & \textbf{36,219} & 5.04 & 8.17 & 41.16 & \textbf{100.0} & 1.42 & \textbf{1.80} \\
\bottomrule
\end{tabular}
\end{table*}

\begin{table*}
\centering
\caption{Per-request service quality metrics (mean $\pm$ standard deviation) at best-performing epochs. Bold values indicate lowest mean wait or lateness.}
\label{tab:sensitivity_request_stats}
\footnotesize
\begin{tabular}{cccccc}
\toprule
\textbf{Battery} & \textbf{Fleet} & \textbf{Load} & \textbf{Wait per Request} & \textbf{Lateness per Request} \\
\textbf{(kWh)} & \textbf{Size} & \textbf{Cap.} & \textbf{Mean $\pm$ Std (s)} & \textbf{Mean $\pm$ Std (s)} \\
\midrule
20 & 4 & 1 & 34.3 $\pm$ 4.8 & \textbf{0.0} $\pm$ 0.0 \\
20 & 4 & 2 & 32.6 $\pm$ 4.9 & 16.5 $\pm$ 5.5 \\
20 & 4 & 3 & 28.9 $\pm$ 4.9 & 30.8 $\pm$ 9.8 \\
20 & 8 & 1 & 52.6 $\pm$ 6.5 & \textbf{0.0} $\pm$ 0.0 \\
20 & 8 & 2 & 47.6 $\pm$ 6.2 & 18.3 $\pm$ 5.8 \\
20 & 8 & 3 & 43.1 $\pm$ 6.2 & 29.5 $\pm$ 8.3 \\
40 & 4 & 1 & 31.4 $\pm$ 4.4 & \textbf{0.0} $\pm$ 0.0 \\
40 & 4 & 2 & \textbf{27.2} $\pm$ 4.4 & 7.2 $\pm$ 3.0 \\
40 & 4 & 3 & 27.7 $\pm$ 4.9 & 16.0 $\pm$ 4.9 \\
40 & 8 & 1 & 49.9 $\pm$ 6.3 & \textbf{0.0} $\pm$ 0.0 \\
40 & 8 & 2 & 45.0 $\pm$ 6.2 & 15.3 $\pm$ 5.3 \\
40 & 8 & 3 & 40.7 $\pm$ 5.7 & 30.1 $\pm$ 7.5 \\
\bottomrule
\end{tabular}
\end{table*}

All configurations demonstrate successful convergence, with the best validation profits achieved between epochs 180 and 265 and final profits ranging from 30{,}824 to 36{,}219. Four-vehicle fleets with 20 kWh batteries require the longest training (189--219 epochs), reflecting challenging optimization under resource constraints. Eight-vehicle fleets converge more rapidly (180--190 epochs), benefiting from a more forgiving optimization landscape when sufficient resources exist. The configuration with 40 kWh batteries, 4 vehicles, and three-passenger capacity requires the longest training (265 epochs), suggesting that combining moderate fleet size with high ride-sharing capacity creates a complex optimization landscape that warrants extended training budgets. Having established convergence across all configurations, we now examine how each parameter affects operational performance.

Fleet size emerges as the dominant factor for service coverage. All 8-vehicle configurations achieve 100\% completion rates regardless of battery size or ride-sharing capacity, whereas 4-vehicle fleets with 20 kWh batteries reach only 85.5--93.2\% completion. Larger fleets deploy 5.04--6.04 vehicles on average (63--76\% utilization), preserving reserve capacity for demand fluctuations, whereas 4-vehicle fleets operate at 90--100\% utilization. This reserve capacity also enables charging flexibility: 8-vehicle fleets with 20 kWh batteries perform 1.52--1.82 charge visits, as some vehicles can charge while others maintain service. Per-vehicle energy consumption is lower in larger fleets (7.60--8.60 kWh for 20 kWh batteries) than in 4-vehicle fleets (9.05--9.55 kWh), indicating better load balancing when service demand is spread more evenly. Total fleet energy consumption is higher for 8-vehicle configurations (41.16--47.43 kWh versus 36.20--50.03 kWh for 4 vehicles), reflecting the additional energy required to achieve complete service coverage. Load factors remain similar across fleet sizes for equivalent ride-sharing capacities.

Battery capacity has the greatest impact on resource-constrained fleets. For 4-vehicle configurations, increasing capacity from 20 to 40 kWh improves completion rates from 85.5--93.2\% to 99.8--100\%, translating into profit improvements of 6--15\%. Larger batteries enable better fleet utilization: 40 kWh configurations require fewer actively deployed vehicles (3.58--3.92, versus 4.00 for all 20 kWh 4-vehicle cases), as individual vehicles can serve more requests per deployment cycle. Energy consumption per vehicle increases substantially with larger batteries (12.76--13.07 kWh versus 9.05--9.55 kWh), reflecting the ability to perform more work without charging interruptions. Notably, 4-vehicle configurations with 20 kWh batteries do not charge (0.00 visits) despite incomplete service coverage. This reflects a learned trade-off under severe capacity constraints: charging would temporarily remove scarce vehicles from service, and the resulting loss in service capacity outweighs the potential benefit of extended range. In contrast, 4-vehicle fleets with 40 kWh batteries charge only when ride-sharing is enabled (3.01 visits for three-passenger capacity), as the combination of larger batteries and ride-sharing creates scenarios in which mid-operation charging becomes beneficial. For 8-vehicle fleets, battery capacity has minimal impact on completion rates since sufficient vehicles already ensure full coverage, though charging patterns differ: 40 kWh configurations require fewer charge visits (0.59--1.65) than 20 kWh configurations (0.00--1.82).

Ride-sharing capacity provides the most substantial benefits in resource-constrained scenarios. For 4-vehicle fleets with 20 kWh batteries, completion rates improve from 85.5\% (single occupancy) to 93.2\% (three passengers), with corresponding profit gains of about 9\%. Load factors increase with capacity (1.30--1.36 for two-passenger and 1.78--1.90 for three-passenger configurations), though these remain below theoretical maxima because of pickup and drop-off sequencing constraints. Ride-sharing also affects energy consumption patterns. Total fleet energy increases slightly with higher ride-sharing (36.20--38.20 kWh for 4-vehicle, 20 kWh configurations) as more requests are served, but for well-resourced configurations the opposite occurs: 8-vehicle fleets with 40 kWh batteries consume less total energy with ride-sharing enabled (41.16 kWh for three passengers versus 47.20 kWh for single occupancy) because fewer vehicle-kilometers are required when passengers share trips. For configurations that already achieve 100\% completion, ride-sharing shows diminishing returns, with profit improvements below 2\%. The highest overall profit (36{,}219) is achieved with 40 kWh batteries, 8 vehicles, and three-passenger capacity, but this represents only a 1.4\% improvement over single occupancy (35{,}715). Wait times and lateness remain stable across all configurations, with mean wait times ranging from 27 to 53 seconds and mean lateness below 31 seconds.

The interactions between these parameters reveal important trade-offs for E-DARP fleet design. Service coverage depends primarily on having sufficient vehicle capacity, achievable through either larger fleets or larger batteries, but the operational characteristics differ substantially. Larger fleets provide redundancy and charging flexibility at the cost of higher total energy consumption and capital investment, whereas larger batteries enable efficient operation with fewer vehicles but create longer charging windows when replenishment is eventually needed. Ride-sharing acts as a multiplier on existing capacity: it substantially improves performance when resources are constrained but offers marginal gains when the fleet is already sufficient. Energy efficiency is maximized through the combination of adequate fleet size and ride-sharing, which together enable lower per-vehicle utilization and shared passenger trips. Charging behavior emerges as a learned strategy that balances service availability against range extension, with the policy avoiding charging when vehicle scarcity makes service interruptions too costly. These results offer practical guidance for fleet operators: compact cities with moderate demand can use smaller fleets with larger batteries to reduce costs, whereas larger cities benefit from more vehicles that allow some to charge while others serve requests. Ride-sharing helps minimize total energy consumption and provides the greatest benefit when resources are limited. The learned policies’ ability to adapt across these configurations highlights their potential as planning tools for fleet sizing and vehicle selection.

\subsection{Stochastic Energy Consumption and Travel Time}
\label{sec:stochastic}

Real-world fleet operations face inherent uncertainty in travel times and energy consumption due to congestion, weather, and driver behaviour. To stress-test policy robustness under such uncertainty, we inject stochasticity directly into the environment during rollout: every time a vehicle commits to an arc $(i,j)$, the actual travel time and energy consumption are drawn from a half-normal distribution whose scale is 10\% of the deterministic cost. Formally, letting $\delta_{ij}$ and $\varepsilon_{ij}$ denote the deterministic estimates, we realize
\begin{align}
\tilde{\delta}_{ij} &= \delta_{ij} + |z|\, (0.1\, \delta_{ij}),\\
\tilde{\varepsilon}_{ij} &= \varepsilon_{ij} + |z|\, (0.1\, \varepsilon_{ij}),
\end{align}
with $z \sim \mathcal{N}(0,1)$. The $|z|$ couples the two samples so that unusually long trips also consume more energy, while the half-normal form guarantees $\tilde{\delta}_{ij} \ge \delta_{ij}$ (vehicles cannot arrive earlier than the deterministic baseline without violating speed limits). Sampling happens online, after the route decision is made, which mirrors deployment: planners commit to deterministic estimates but discover the actual delay only after traversing the edge.

We treat vehicles sequentially in a rolling re-plan scheme. After a vehicle completes its stochastic route, the dispatcher observes the realized energy consumption and travel times, quantities that were uncertain at the time of assignment, along with which pickups and deliveries were actually served. The next vehicle is then re-optimised using this revealed information. In effect, later vehicles benefit from a form of look-ahead: uncertainty that would still be unresolved in a fully simultaneous dispatch has already materialised by the time their routes are planned. This sequential revelation of stochastic outcomes provides an optimistic benchmark, yet it mirrors practical operations where new assignments are issued only after previous trips finish and is consistent with re-optimisation workflows when time windows are tight.

To study whether explicit exposure to uncertainty helps, we compare two training regimes: one where the model is trained purely on deterministic data, and one where the half-normal sampling is enabled throughout training so every trajectory experiences online noise. Both policies are evaluated on 2000 stochastic validation instances, with results reported in Table \ref{tab:stochastic_results}.

\begin{table}[t]
\caption{Performance comparison under stochastic evaluation (10\% half-normal noise). Both models are evaluated on 2000 instances with stochastic travel times and energy consumption.}
\label{tab:stochastic_results}
\centering
\begin{tabular}{lcc}
\hline
\textbf{Metric} & \textbf{Deterministic Training} & \textbf{Stochastic Training} \\ \hline
\multicolumn{3}{l}{\textit{Primary Metrics}} \\
Average Reward & 17.94 & 17.94 \\
Completion Rate (\%) & 100.00 & 100.00 \\
Total Waiting Time & 0.063 & 0.065 \\
Total lateness & 0.022 & 0.023 \\ \hline
\multicolumn{3}{l}{\textit{Fleet Utilization}} \\
Vehicles Used & 6.45 & 6.63 \\
Charging Visits & 1.50 & 1.11 \\
Low SoC Blocks & 4.07 & 2.65 \\
Fleet Exhausted (\%) & 4.8 & 11.0 \\ \hline
\multicolumn{3}{l}{\textit{Energy Statistics}} \\
Mean SoC & 0.766 & 0.773 \\
Energy Used & 2.62 & 2.61 \\ \hline
\end{tabular}
\end{table}

The results reveal that both training regimes achieve nearly identical performance under stochastic evaluation. The deterministic-trained policy attains an average reward of 17.94 with a completion rate of 100.00\%, matching the stochastic-trained policy. This suggests that deterministic training already produces policies that are robust to moderate (10\%) variability in travel times and energy consumption.

Interestingly, the two policies exhibit different operational strategies. The stochastic-trained model uses slightly more vehicles on average (6.63 vs 6.45) but makes fewer charging visits (1.11 vs 1.50). This indicates that exposure to uncertainty during training encourages more conservative battery management: the policy learns to maintain higher state of charge levels to buffer against unexpected energy consumption, reducing the need for en-route charging. This is further evidenced by the lower frequency of low SoC blocking events (2.65 vs 4.07) and slightly higher mean SoC (0.773 vs 0.766) for the stochastic-trained model.

However, the more conservative strategy comes with a trade-off: the stochastic-trained policy exhausts the vehicle fleet more frequently (11.0\% vs 4.8\% of instances). By deploying more vehicles with fewer charging stops, the policy occasionally runs out of fleet capacity before completing all requests, though this does not significantly impact the overall completion rate due to the sequential re-planning scheme.

These findings suggest that for moderate levels of uncertainty, deterministic training provides sufficient robustness. The learned policies generalize well to stochastic conditions without requiring explicit exposure to noise during training. This has practical implications: operators can train models on deterministic historical data and still expect reliable performance under real-world variability.

\section{Conclusion}
\label{sec:conclusion}

This paper presents a deep reinforcement learning framework for the Electric Dial-a-Ride Problem (E-DARP), addressing the operational challenges of autonomous electric vehicle fleets serving on-demand mobility requests. The GREAT encoder architecture proves particularly well-suited for E-DARP by operating directly on edges rather than deriving edge information from node coordinates. This design naturally handles the non-Euclidean nature of energy-aware routing, where edge features such as travel times and energy consumption depend on factors like road gradient and traffic conditions that cannot be captured by coordinate-based distance calculations. The edge-centric formulation enables the policy to represent asymmetric costs, asymmetric real world energy consumption, and asymmetric graph structures.

Benchmark experiments against exact methods demonstrate that the learned policy achieves near-optimal solutions, with 0.00\% optimality gaps on small instances (16 requests) and 0.40\% gaps on larger instances (50 requests). The framework provides computational speedups of 21 to 44 times for small instances and over 7,200 times for instances where exact methods approach their time limits.

Curriculum learning enables scaling to 250-request instances, where the policy outperforms ALNS by 9.5\% in solution quality while achieving 100\% completion and 774 times faster computation. The learned policy exhibits emergent ridesharing behavior with 3.5 times higher vehicle utilization than ALNS, demonstrating effective multi-passenger coordination.

Hyperparameter sensitivity analysis reveals that smaller transformer architectures (hidden dimension 64--128) outperform larger models while exhibiting stable convergence, enabling efficient deployment without sacrificing solution quality. The reward function analysis confirms that adequate completion weight is essential for achieving full service coverage.

Sensitivity analysis across twelve configurations on 100-request scenarios reveals critical operational trade-offs and demonstrates the framework's utility as a fleet planning tool. Battery capacity emerges as the dominant factor for resource-constrained fleets, with larger batteries enabling near-complete service coverage. Fleet size proves critical for service reliability, with larger fleets achieving full completion while maintaining reserve capacity for demand fluctuations. Ride-sharing provides the greatest benefit when resources are limited but shows diminishing returns when fleet capacity is sufficient. Service quality metrics remain stable across all configurations, with mean wait times below one minute and minimal lateness. These results illustrate how the framework can guide fleet sizing and vehicle selection decisions for different urban contexts.

Several limitations suggest directions for future research. The framework assumes deterministic demand arrival patterns, whereas real-world operations face uncertain request arrivals, customer no-shows, and cancellations. Extending the methodology to handle dynamic demand through online learning or anticipatory policies would enhance operational reliability. Additionally, the formulation assumes a homogeneous fleet where all vehicles share identical battery capacity, passenger capacity, and energy consumption characteristics. Real-world fleets often comprise mixed vehicle types with varying ranges and capacities, requiring extensions to handle heterogeneous vehicle assignment and routing decisions. Furthermore, the current framework discretizes charging decisions by fixing charging duration at each station visit, whereas exact formulations optimize continuous charging times. This discretization may contribute to the small optimality gaps observed on benchmark instances (0.4\% for 50-request problems) and could limit performance in scenarios requiring precise energy management. Future work could explore hybrid approaches that combine learned routing policies with optimization-based charging decisions, or action spaces that include multiple discrete charging duration options.

This research demonstrates that edge-based deep reinforcement learning can effectively address complex vehicle routing problems with integrated operational decisions, achieving near-optimal performance with computational efficiency suitable for real-time deployment. The inference time of approximately 10 seconds enables practical deployment in operational environments where traditional optimization methods requiring hours become computationally intractable. Beyond empirical performance, the edge-centric formulation offers a more structured representation of routing decisions by operating directly on graph edges rather than latent node embeddings. This added structural alignment with the underlying combinatorial problem may facilitate future theoretical analysis of learning dynamics, stability, and convergence properties of routing policies. While formal convergence guarantees for deep reinforcement learning remain an open challenge, the proposed architecture provides a promising foundation for such analysis. As cities worldwide pursue electrification and autonomous mobility services, the framework presented here provides both algorithmic advances and practical guidance for fleet deployment decisions.

\section{Acknowledgement}
This work was co-funded by Vinnova, Sweden through the project: Simulation, analysis and modeling of future efficient traffic systems. This work was in part supported by the Transport Area of Advance within Chalmers University of Technology. The computations were enabled by resources provided
by the National Academic Infrastructure for Supercomputing
in Sweden (NAISS) at Chalmers e-Commons partially funded
by the Swedish Research Council through grant agreement no.
2022-06725.

\printcredits
\newpage
\bibliography{cas-refs}




\end{document}